\documentclass[longauth]{aa} 

\usepackage{txfonts}
\usepackage{natbib}
\usepackage{graphicx}
\usepackage{hyperref}
\usepackage{multirow}
\usepackage{longtable}
\usepackage{xcolor}
\newcommand{\grb}{GRB\,171205A}
\newcommand{\hi}{\sc Hi}
\newcommand{\zhi}{\mbox{$z_{\rm HI}$}}
\newcommand{\mhi}{\mbox{$M_{\rm HI}$}}
\newcommand{\lphi}{\mbox{$L'_{\rm HI}$}}
\newcommand{\mhtwo}{\mbox{$M_{\rm H2}$}}
\newcommand{\kms}{\mbox{km\,s$^{-1}$}}

\newcommand{\msun}{\mbox{$M_\odot$}}

\begin{document}

   \title{HI and CO spectroscopy of the unusual host of GRB\,171205A}

   \subtitle{A grand design spiral galaxy with a distorted HI field}

   \author{A. de Ugarte Postigo \inst{1,2}
          \and
          M.~Micha\l{}owski, \inst{3}
          \and
          C.~C.~Th\"one, \inst{4}
          \and 
          S.~Martin, \inst{5,6}
          \and
          A.~Ashok, \inst{7,8,9}
          \and
          J.~F.~Ag\"u\'i~Fern\'andez \inst{10}
          \and
          M.~Bremer \inst{11}
          \and
          K.~Misra \inst{12}
          \and
          D.~A.~Perley \inst{13}
          \and
          K.~E.~Heintz \inst{14}
          \and          
          S.~V.~Cherukuri \inst{7}
          \and
          W.~Dimitrov \inst{3}
          \and 
          T. G\'eron \inst{15}
          \and
          A.~Ghosh \inst{12}
          \and
          L.~Izzo \inst{16,17} 
          \and
          D.~A.~Kann \inst{18} 
          \and
          M.~P.~Koprowski \inst{19}
          \and
          A.~Le\'sniewska \inst{3}
          \and
          J. K. Leung \inst{20,15,21}
          \and          
          A.~Levan \inst{22,23}
          \and
          A.~Omar \inst{12,24}
          \and
          D.~Oszkiewicz\inst{3}
          \and
          M.~Poli\'nska \inst{4}
          \and
          L.~Resmi \inst{7}
          \and
          S.~Schulze \inst{25}
          }
   \institute{
    Observatoire de la C\^ote d'Azur, Universit\'e C\^ote d'Azur, Artemis Boulevard de l'Observatoire, 06304 Nice, France
    \and
    Aix Marseille Univ, CNRS, CNES, LAM Marseille, France
   \and
   Astronomical Observatory Institute, Faculty of Physics, Adam Mickiewicz University, ul. S\l{}oneczna 36, 60-286 Pozna\'n, Poland
   \and 
   Astronomical Institute, Czech Academy of Sciences, Fri\v cova 298, Ond\v rejov, Czech Republic
   \and
   European Southern Observatory, Alonso de C\'ordova, 3107, Vitacura, Santiago 763-0355, Chile
   \and
   Joint ALMA Observatory, Alonso de C\'ordova, 3107, Vitacura, Santiago 763-0355, Chile
   \and
   Department of Earth \& Space Sciences, Indian Institute of Space Science \& Technology, Trivandrum 695547, India 
   \and
   Max Planck Institute for Gravitational Physics (Albert Einstein Institute), D-30167 Hannover, Germany 
   \and
   Leibniz Universit\"at Hannover, D-30167 Hannover, Germany 
   \and
   Instituto de Astrof\'isica de Andaluc\'ia - CSIC, Glorieta de la Astronom\'ia s/n, 18008 Granada, Spain
   \and
   Institut de Radioastronomie Millim\'etrique (IRAM), 300 Rue de la Piscine, F-38406 Saint Martin d'H\`eres, France
   \and
   Aryabhatta Research Institute of observational sciencES (ARIES), Manora Peak, Nainital 263001, India 
   \and
   Astrophysics Research Institute, Liverpool John Moores University, Liverpool Science Park, 146 Brownlow Hill, Liverpool L3 5RF, UK 
   \and
   Cosmic Dawn Center (DAWN), Niels Bohr Institute, University of Copenhagen, Jagtvej 128, DK-2200 Copenhagen N, Denmark
   \and
   Dunlap Institute for Astronomy \& Astrophysics, University of Toronto, 50 St. George Street Toronto, ON M5S 3H4, Canada
   \and
   Dark Cosmology Centre (DARK), Niels Bohr Institute, University of Copenhagen, Jagtvej 128, 2200 Copenhagen N, Denmark
   \and
   INAF-Osservatorio Astronomico di Capodimonte, Salita Moiariello 16, 80131, Napoli, Italy
   \and
   Hessian Research Cluster ELEMENTS, Giersch Science Center, Max-von-Laue-Straße 12, Goethe University Frankfurt, Campus Riedberg, D-60438 Frankfurt am Main, Germany
   \and
   Institute of Astronomy, Faculty of Physics, Astronomy and Informatics, Nicolaus Copernicus University, Grudziadzka 5, 87-100 Torun, Poland
   \and
   David A. Dunlap Department of Astronomy \& Astrophysics, University of Toronto, 50 St. George St., Toronto, Ontario, M5S 3H4, Canada
   \and
   Racah Institute of Physics, The Hebrew University of Jerusalem, Jerusalem, 91904, Israel
   \and
   Department of Astrophysics/IMAPP, Radboud University,  6525 AJ Nijmegen, The Netherlands
   \and
   Department of Physics, University of Warwick, Coventry, CV4 7AL, UK
   \and
   SPASE, Indian Institute of Technology, Kanpur 208016 India 
   \and
   The Oskar Klein Centre, Physics Department of Physics, Stockholm University, Albanova University Center, SE 106 91 Stockholm, Sweden 
 }

   \date{Received ; accepted }

 
  \abstract
   {Long gamma-ray bursts are produced by the collapse of very massive stars after a very short life time. They are found very close to the place in which they formed, normally near the most prominent star-forming regions of star-forming galaxies.}
   {GRB\,171205A is the fourth closest known GRB and happened in the outskirts of a grand-design spiral galaxy, a peculiar location within an atypical GRB host. In this paper we present a highly-resolved study of the molecular gas of this GRB host, with observations performed by ALMA at the CO($1 - 0$) transition to investigate the environment in which this burst happened. These data are compared with GMRT atomic {\hi} observations, and complemented with data at other wavelengths to provide a broad-band view of the galaxy.}
   {The ALMA observations were performed with a spatial resolution of $0.2^{\prime\prime}$ and a spectral resolution of 10 km/s at the time when the afterglow had a flux density of $\sim$53 mJy. This allowed us to attempt a molecular study both in emission and absorption. We used the {\hi} observations to study both the host galaxy and its extended environment.}
   {The CO emission shows an undisturbed grand-design spiral structure with a central bar, and no significant emission at the location of the GRB. The line-of-sight study towards the GRB does not reveal any CO absorption down to a column density limit of $<10^{15}$ cm$^{-2}$. This argues against the hypothesis of the progenitor forming within a massive molecular cloud. The molecular gas is distributed along the galaxy arms with higher concentration in the regions dominated by dust. However, the {\hi} gas distribution does not follow the same pattern as the stellar light or the one of molecular gas and is concentrated in two blobs and with no emission towards the centre of the galaxy. Overall the {\hi} is slightly displaced towards the southwest of the galaxy, where the GRB exploded. Within the extended neighbourhood of the host galaxy, we identify another prominent {\hi} source at the same redshift, at a projected distance of 188 kpc, which we confirm with optical spectroscopy.}
   {Our observations show that the progenitor of this GRB is not associated to a massive molecular cloud, but more likely related to low-metallicity atomic gas. The distortion in the {\hi} gas field is the strongest indicator of an odd environment that could have triggered a star formation episode and could be linked to a past interaction with the companion galaxy.}

   \keywords{gamma-ray burst: individual: GRB\,171205A; galaxies: ISM; submillimeter: galaxies}
   \maketitle
%
\section{Introduction}

The majority of gamma-ray bursts (GRBs) are the so-called long-gamma-ray bursts (LGRBs, with duration $>2$ seconds), which are normally produced during the collapse of short-lived, very massive stars \citep[e.g.][]{Woosley2006ARAA,Hjorth2012Book,Cano2017AdAst}. Because of this, LGRBs are found in star forming galaxies \citep{Paczynski1998ApJ} and, in particular, within or near some of the most active star-forming regions of those host galaxies, in which their massive progenitors are born \citep{Fruchter06}. A second, somewhat less common and less luminous type are short GRBs (SGRBs, duration $<2$ s) which are associated with the merger of two neutron stars and have been linked to the emission of gravitational waves \citep{Abbott2017ApJLIGOGBM,Abbott2017ApJMMA}. Recent results have demonstrated that duration is far from a direct map to progenitor type \citep[e.g.][]{Rastinejad2022,Troja2022,Levan2023}. However, within this paper, we will be discussing solely LGRBs where strong evidence ties them to stellar core collapse. 

LGRBs are some of the most luminous sources seen in the Universe, with gamma-ray energy releases of the order of $10^{51}$ erg in just a few seconds \citep{Kulkarni1999Nature}. This high-energy emission is followed by an afterglow that emits at all electromagnetic wavelengths and can be very luminous for a short period of time \citep{Akerlof1999Nature,Kann2007AJ,Racusin2008Nature,Bloom2009ApJ}, after which it gradually fades away. The spectrum of the afterglow is normally well described by a synchrotron emission generated by the interaction of an ultra-relativistic jet with the interstellar medium that surrounds the progenitor. This spectrum evolves with time from high to low energies as the shock decelerates \citep{Sari1998ApJ}. This means that the optical peak is often observed within tens of seconds of the burst onset \citep{Molinari2007AA}, while it can take hours to peak at submillimetre wavelengths \citep{deugarte2012} and days at decimetre wavelengths \citep{Chandra2012}.

The luminosity of LGRB afterglows, together with their simple synchrotron spectrum, make them useful beacons to study their local and intervening environment in absorption \citep[e.g.][]{Vreeswijk2007AA,Hartoog2015AA,deugarte2018}, shining from the heart of the star forming regions in which GRBs are produced. On the other hand, once localised through the GRB emission, we can also study these star-forming host galaxies in emission using different techniques \citep{Hjorth2012,Perley2016A,Michalowski18grb,Hatsukade2020,Thoene2021,Schady2023}. This is a unique feature of GRB hosts which is not accessible with almost any other known astronomical source.

Within LGRBs there is a subclass known as low-luminosity GRBs (llGRBs) that can be between 10 and 1000 times fainter than typical LGRBs \citep{Soderberg2004Nature,Soderberg2006Nature,Sun2015}. Due to this lower luminosity, we normally only detect them at low redshift. However, they are also over 100 times more common \citep{Pescalli2015}, which allows us to observe many more GRBs in the local Universe than if only the luminous class of LGRBs existed. Indeed, most of the GRBs that we detect at low redshift are llGRBs. These sources typically have a softer spectrum than the average LGRB and are often characterised by a single-peak prompt emission light curve \citep[e.g. GRB 060218 and 100316D,][]{Campana2006Nature,Starling2011MNRAS}.

GRB\,171205A was a low-luminosity event located in the outskirts of a grand-design spiral host galaxy at a redshift of $z=0.0368$ \citep{Izzo19Nat}, a rather atypical environment for a LGRB. It is the fourth closest LGRB known to date, after GRB\,980425, at a redshift of $z=0.0085$ \citep{Galama1998Nature}, GRB\,111005A, at $z=0.0133$ \citep{Michalowski18grb,Tanga2018AA}, and GRB\,060218, at $z=0.0331$ \citep{Pian2006Nature}, all of which were also low-luminosity events. It was associated with SN 2017iuk \citealp{deUgarte2017b}, a broad-lined Ic supernovae with an early cocoon emission \citep{Izzo19Nat}, which unambigously links this event to the death of a massive star. Although it did not have a luminous afterglow, GRB\,171205A was one of the brightest events ever observed thanks to its proximity, being the second brightest afterglow observed at millimetre wavelengths \citep{deugarte2017}, only surpassed by GRB\,030329 \citep{Smith2005AA}. This bright afterglow motivated us to attempt spectroscopic observations in search of molecular absorption.
 
In this paper we present a study of the host galaxy environment of GRB\,171205A, mainly based on decimetre and millimetre observations tuned to the {\hi} and CO($1-0$) transitions, respectively. We analyse the host emission in these transitions and discuss our search for molecular absorption in the data. Section 2 introduces the observations which are analysed in Sect. 3. In Sect. 4 we discuss our results and Sect. 5 presents our conclusions. Throughout the paper we use a flat $\Lambda$CDM cosmology with $\Omega_m=$0.286, $\Omega_\Lambda=$0.714 and H$_0=$69.6 \cite{Bennett2014}. 

\section{Observations}

GRB\,171205A triggered the Burst Alert Telescope (BAT) onboard the \textit{Neil Gehrels Swift Observatory} \cite[][{\it Swift} hereafter]{Gehrels2004} at 07:20:43 UT on 5 December 2017 and was rapidly localised by the X-ray Telescope (XRT) within an uncertainty of a few arcseconds \citep{Delia2017}. This allowed a rapid identification of its likely host galaxy as a nearby grand-design spiral \citep{Izzo2017a}. Subsequent optical and near-infrared (nIR) observations from VLT identified the afterglow, confirmed the association of the afterglow and host galaxy, and determined their redshift to be $z = 0.0368$ \citep{izzo2017b}. The burst had a T$_{90}$ duration of $189.4\pm35.0$~s, which classified it as a LGRB, but its isotropic equivalent energy release in gamma-rays was only $2.68\times10^{49}$ erg (based on high-energy data from \citealt{Tsvetkova2021ApJ}), making it a llGRBs. The discovery of such a nearby event triggered a broad follow-up campaign involving observations by numerous observatories and at almost all observable bands.

We performed follow-up observations with the NOEMA observatory that started 20.3 hr after the burst onset \citep{deugarte2017}. The observations consisted of continuum observations at 90 and 150~GHz that will be studied, together with other continuum observations, in a separate work (de Ugarte Postigo in prep.). Using these data we identified the afterglow at a flux density of 35 mJy, already the second brightest ever detected. This motivated the trigger of our ALMA target of opportunity programme to search for molecular absorption along GRB lines-of-sight.

\subsection{ALMA CO($1-0$) imaging and spectroscopy}

ALMA Band 3\footnote{Project ADS/JAO.ALMA\#2017.1.01695.T, P.I. A. de Ugarte Postigo} was set up to target the CO($1-0$) transition at the redshift $z=0.0368$ (i.e. 111.180~GHz) and phase centred at $\alpha_{J2000}=1$1:09:39.52, $\delta_{J2000}=-1$2:35:18.34. The correlator was configured with four spectral windows, one of them centred at the frequency of the redshifted CO($1-0$). The bandwidth of 1.875~GHz was sampled at an original frequency resolution 0.977~MHz ($\sim2.6$ km~s$^{-1}$). The other three coarse resolution bands were arranged for continuum detection and will be published in a separate photometric study (de Ugarte Postigo in prep.).

   \begin{figure}
   \centering
   \includegraphics[width=\hsize]{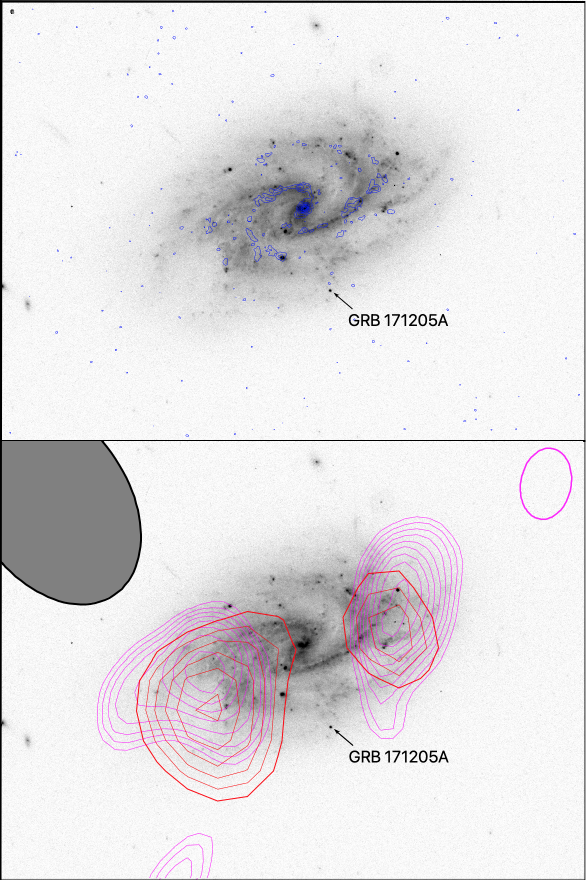}
      \caption{ALMA CO (top, in blue) and GMRT {\hi} (bottom, in red) emission contours of the host galaxy as compared to a visible light (filter $F606W$) image from HST. The field of view is $48^{\prime\prime}\times36^{\prime\prime}$, North is to the top and East to the left. In the bottom panel we have included in magenta, for comparison, the JVLA contours from the reduction done by \citep{Thoene2023}. The contours start at $3-\sigma$ with $2-\sigma$ increments in CO, and in $1-\sigma$ increments in HI. The ellipse at the top left of each figure indicates the resolution of each observation. The magenta ellipse corresponds to the spatial resolution of the JVLA data. While the CO molecular gas traces the bar and spiral arms of the galaxy, and in particular some of the dust features within the arms (shown as lighter patches in the HST image), the {\hi} neutral gas emission does not follow the distribution of stellar mass and shows a double blob structure slightly offset towards the southwest.
              }
         \label{figure:almahost}
   \end{figure}

The observations were split over two days, with 6 consecutive executions on December 7th, 2017. The last execution on December 7th was of poor quality and did not pass the ALMA quality check, so it was repeated on the 8th of December. We combined a total integration time of 4.9 hours on source. In all executions, band pass, absolute flux, and pointing were calibrated with respect to the quasar J1058+0133, while complex gain calibration was obtained using quasar J1127$-$1857 as reference.

The data were calibrated and imaged with CASA \citep[v5.1.1; ][]{McMullin2007}. Thanks to the relatively bright continuum detection at the GRB position during these observations ($>50$~mJy in all individual executions), the individual data were amplitude and phase self-calibrated. However, the improvement was marginal as expected in low frequency ALMA observations. The data were smoothed to a velocity resolution of $\sim10$ km~s$^{-1}$. At this resolution we achieved an r.m.s. noise of 0.2~mJy.

The spatial resolution was $0\farcs31\times0\farcs24$ (equivalent to $228\times176$ pc at the redshift of the galaxy) at a position angle of $-78^\circ$. 
A map of the continuum was obtained by aggregating the continuum spectral windows from all six executions. Apart from the continuum at the GRB position, no continuum is detected towards the host galaxy centre down to an r.m.s. of 15~$\mu$Jy, slightly worse than the limit obtained in a later observation, when the afterglow was less prominent (see section 2.3.3).

From this data set we were able to extract an afterglow spectrum covering the CO($1-0$) transition with a signal to noise ratio (SNR) of $\sim270$, way beyond other published attempts \citep{deugarte2018,deugarte2020,Michalowski2018co}.

Additionally, we produced a continuum subtracted data cube that allows us to study the host galaxy in CO($1-0$) emission. Continuum was subtracted in the uv plane with the CO line emission free channels.

\subsection{GMRT {\hi} imaging and spectroscopy}

To complement our observations of the host galaxy in CO($1-0$), we observed the field of {\grb} with the Giant Metrewave Radio Telescope (GMRT) in search of {\hi} emission. We obtained 12\,hr of observations split between two epochs on 2018-02-11 and 2018-03-15\footnote{Project no.~ddtB305, PI: M.~Micha{\l}owski}. The objects 3C147 and 3C286 were used as primary calibrators, whereas object 1130-148 was used as a secondary calibrator. The observing frequency was 1.362\,GHz, covering the {\hi} line. The total bandwidth was 16.7\,MHz, and the channel width was 32.6\,kHz.

We reduced the data in the Astronomical Image Processing System ({\sc Aips}; \citealt{vanmoorsel96}). The data were flagged following standard procedures.
The {\hi} data from the two epochs were combined in AIPS and processed together. The {\hi} cubes were produced by Fourier transforming the data in each channel individually and then subtracting the possible continuum emission. We tapered the data at 25 and $10\,\mbox{k}\lambda$ and obtained a beam size of $13\arcsec\times19\arcsec$ and $21\arcsec\times27\arcsec$, respectively. These data reveal a prominent emission at the position of the host galaxy of GRB\,171205A, although not quite coincident with the centroid of the optical or CO emission. At similar velocity we detect further {\hi} emission throughout the field of view of the cube, including a high-significance one that matches a galaxy detected in optical surveys, which we refer to as \textit{companion galaxy} throughout the paper (see Sect. 3.5 for further details).

A similar {\hi} dataset, with higher spatial resolution, obtained at JVLA, has been recently published by \citep{arabsalmani2022} and later revisited by \citep{Thoene2023}. Throughout the paper we use the latter reduction for comparison to our results.

\subsection{Complementary observations}

In this section we present complementary observations that allow us to place into context the data described in the previous sections. This includes optical afterglow spectroscopy, high spatial-resolution imaging from the Hubble Space Telescope (HST), further continuum radio observations of the host, and archival host galaxy data from infrared to ultraviolet. We also include imaging and spectroscopic observations of a nearby companion galaxy.

\subsubsection{GTC spectroscopy}

As a reference of the line-of-sight environment from optical data, we use optical spectroscopy obtained almost simultaneously to the CO spectrum with the OSIRIS instrument\footnote{Project  GTCMULTIPLE2J-17B, PI: C.~C.~Th\"one}, mounted on the 10.4 m GTC telescope, which was already presented by \cite{Izzo19Nat}. At the time of the observation, the emission was dominated by the cocoon emission discovered in that paper. The goal here is to use the optical light of the source as a back-illuminating source to search for absorption features in a similar way as we do with the millimetre data.

Of the strong features usually found in GRB afterglows \citep{deugarte2012}, only the CaII $\lambda 3935$ and CaII $\lambda 3970$ lines are covered by the spectroscopy at this redshift. These CaII absorptions are undetected down to a rest-frame equivalent width limit of 0.9 {\AA}. The typical values observed for these features are 1.29 and 0.93 {\AA}, respectively, meaning that these absorption features are weaker than average, although with only a weak limit. This implies that the line-of-sight does not encounter large amounts of material. However, there is a marginal detection of NaID$_1$ at $0.34\pm0.12$ {\AA}, which implies a non-negligible column density of metals.

\subsubsection{HST imaging} 

We use HST imaging observations of the host galaxy that allow us to do a comparison with the ALMA data at a comparable spatial resolution. These observations are presented by \citet{Thoene2023} and have a spatial resolution of 0\farcs10. In this paper we compare our data with HST images obtained with the $F606W$ filter on 2 July 2018, when the afterglow was still faintly detected. The $F606W$ at the redshift of GRB\,171205A includes the H-$\alpha$ emission, which can be used as a tracer of star formation. Unlike what we often see in other GRB sites, neither the location of the optical afterglow, nor its surroundings seem to correspond to a prominent star forming region.

\subsubsection{Host galaxy continuum observations at submillimetre, millimetre and decimetre wavelengths}

Our long-term follow-up effort included continuum observations performed with the ALMA observatory. In this paper we use some of the data at millimetre and submillimetre wavelengths obtained within the context of these programmes to search for continuum emission from the host galaxy.

The core of the host galaxy is detected in the observation performed on 18 April 2018 at 343.5 GHz, with a flux density of $371\pm81$ $\mu$Jy. This observation was aimed at detecting the afterglow\footnote{Project 2017.1.01526.T, PI: D.~A.~Perley} and covers the core of the galaxy, but not the northern part of the spiral structure.  In any case, the galaxy structure beyond the core is too faint to be detected. To estimate the flux density of the complete galaxy we assume that the ratio between the flux density of the core and that of the complete galaxy is similar to that of the CO image. This is supported by the fact that CO is shown in Fig. 1 to be a good tracer of dust, and the millimetre continuum emission is dominated cold dust, but we note that it has an inherent uncertainty that is difficult to quantify. By using this approximation, we derive a multiplying factor of $\sim 2.7$, which we apply to our measurements and determine a host galaxy flux density at 343.5 GHz of $1002\pm220$ $\mu$Jy. In a similar way, we look at the deepest available image in the 97.5 GHz band, obtained on 11 October 2018\footnote{Project  2018.1.01635.S, PI: C.C.~Th\"one}. In this case, there is no detection at the position of the core down to a $3\sigma$ limit of 39 $\mu$Jy. Applying the same factor, we estimate the limit to be $< 105\ \mu$Jy for the complete galaxy. This host photometry is presented in Table~\ref{table:hostphot}.

During the observation of HI obtained with GMRT, the afterglow still had a flux density of $\sim4$ mJy, and it contaminated a possible continuum detection of the host galaxy. However, in our continued monitoring we obtained observations in March and June 2022\footnote{Project no. 42\_110, PI: K. Misra}, more than 5 years after the burst. In these continuum images the afterglow, although still detected, has significantly decayed and the core of the host galaxy can be easily separated. We measure an avegerage flux density from the two epochs of $150\pm50\ \mu$Jy at a frequency of 1.39 GHz. Applying the same correction factor mentioned before, we estimate a total flux density for the complete host galaxy of $405\pm135\ \mu$Jy. Again, we caution the reader about the uncertainty of this approximation.

\begin{table}
\caption{Host galaxy photometry. The magnitudes are in the AB system and have been corrected for Galactic extinction is E($B-V$) $ = 0.045$ mag, or A$_V= 0.138$ mag \citep{Schlafly2011}.}       
\label{table:hostphot}      
\centering                         
\small{
\begin{tabular}{c c c c}
\hline\hline
Band            & $\lambda$     & AB Mag            & Flux density      \\
                & ($\mu m$)     &                   & ($mJy$)             \\
\hline
GMRT 1.39 GHz   &$2.16\times10^{5}$& $17.4\pm0.3$   &  $0.405\pm0.135$  \\
ALMA B3         &  3074.8       & $>18.85$          &  $<0.105$         \\
ALMA B6         &  872.8        & $16.40\pm0.25$    & $1.00\pm0.22$     \\
WISE4           &  22.0883      & $13.71\pm0.10$    & $11.9\pm1.2$      \\
WISE3           &  11.5608      & $14.10\pm0.05$    & $8.3\pm0.4$       \\ 
WISE2           &  4.6028       & $15.63\pm0.05$    & $2.03\pm0.10$     \\ 
WISE1           &  3.3526       & $15.14\pm0.04$    & $3.19\pm0.12$     \\
VISTA $K_S$     &  2.1470       & $14.535\pm0.009$  & $5.65\pm0.05$     \\
VISTA $H$       &  1.6460       & $14.361\pm0.022$  & $7.05\pm0.09$     \\
VISTA $J$       &  1.2520       & $14.289\pm0.013$  & $6.54\pm0.13$     \\
VISTA $Y$       &  1.0200       & $14.250\pm0.009$  & $5.57\pm0.05$     \\
Pan-STARRS $y$  &  0.96135      & $14.50\pm0.02$    & $5.75\pm0.11$     \\ 
Pan-STARRS $z^\prime$ & 0.86686 & $14.72\pm0.06$    & $4.70\pm0.26$     \\ 
Pan-STARRS $i^\prime$ & 0.75037 & $14.83\pm0.03$    & $4.25\pm0.12$     \\ 
Pan-STARRS $r^\prime$ & 0.61564 & $15.11\pm0.03$    & $3.28\pm0.09$     \\ 
UVOT $v$        &  0.54114      & $15.47\pm0.04$    & $2.36\pm0.08$     \\ 
Pan-STARRS $g^\prime$ & 0.48109 & $15.74\pm0.02$    & $1.84\pm0.04$     \\
UVOT $b$        &  0.43463      & $15.85\pm0.02$    & $1.66\pm0.03$     \\ 
UVOT $u$        &  0.35210      & $16.92\pm0.03$    & $0.619\pm0.017$   \\ 
UVOT $uvw1$     &  0.26841      & $17.36\pm0.03$    & $0.413\pm0.012$   \\
GALEX $NUV$     &  0.23047      & $17.77\pm0.03$    & $0.283\pm0.008$   \\ 
UVOT $uvm2$     &  0.22458      & $17.64\pm0.02$    & $0.319\pm0.006$   \\ 
UVOT $uvw2$     &  0.20857      & $17.62\pm0.02$    & $0.325\pm0.006$   \\
GALEX $FUV$     &  0.15490      & $18.40\pm0.09$    & $0.158\pm0.013$   \\
\hline
\end{tabular}}
\end{table}
\begin{table}
\caption{Photometry of the companion galaxy of the GRB\,171205A host galaxy. The magnitudes are in the AB system and have been corrected for Galactic extinction \citep{Schlafly2011}.}
\label{table:hostphotcomp}
\centering
\small{
\begin{tabular}{c c c c}
\hline\hline
Band            & $\lambda$     & AB Mag            & Flux density      \\ 
                & ($\mu m$)     &                   & ($mJy$)           \\ 
\hline 
WISE3           &  11.5608      & $17.049\pm0.042$  & $0.55\pm0.02$     \\
WISE2           &  4.6028       & $18.103\pm0.084$  & $0.21\pm0.02$     \\
WISE1           &  3.3526       & $17.391\pm0.076$  & $0.40\pm0.03$     \\
VISTA $K_S$     &  2.1470       & $16.471\pm0.050$  & $0.94\pm0.04$     \\
VISTA $H$       &  1.6460       & $16.314\pm0.035$  & $1.08\pm0.03$     \\
VISTA $J$       &  1.2520       & $16.407\pm0.033$  & $0.99\pm0.03$     \\
VISTA $Y$       &  1.0200       & $16.524\pm0.022$  & $0.89\pm0.02$     \\
PanSTARRS $y$   &  0.96135      & $16.499\pm0.076$  & $0.91\pm0.06$     \\ 
PanSTARRS $z^\prime$ &  0.86686 & $16.740\pm0.057$  & $0.73\pm0.04$     \\
PanSTARRS $i^\prime$ &  0.75037 & $16.727\pm0.040$  & $0.74\pm0.03$     \\ 
PanSTARRS $r^\prime$ &  0.61564 & $16.928\pm0.021$  & $0.61\pm0.01$     \\
PanSTARRS $g^\prime$ &  0.48109 & $17.235\pm0.050$  & $0.46\pm0.02$     \\
GALEX $NUV$   &  0.23047        & $20.344\pm0.050$  & $0.026\pm0.001$   \\
GALEX $FUV$   &  0.15490        & $21.367\pm0.180$  & $0.010\pm0.002$   \\
\hline
\end{tabular}}
\end{table}

\subsubsection{Archival host galaxy data}
\label{hostphot}

Due to the size and brightness of the host galaxy of GRB\,171205A, it had already been observed by numerous survey telescopes at different wavelengths and included in their catalogues. In this section we compile these observations. Whenever possible, we refined the available photometry using consistent apertures with a radius of 25\arcsec. The observations include mid-infrared data from the WISE observatory \citep{Cutri2013a}, near-infrared data from the VISTA Hemisphere Survey \citep[VHS,][]{McMahon2013} optical data from PanSTARRS \citep{Flewelling2016}, optical and ultraviolet (UV) data from UVOT \citep{Roming2005SSRv} and further UV data from \textit{GALEX} \citep{Bianchi2011}. The photometry collected from these observatories, corrected for Galactic extinction, is presented in Table~\ref{table:hostphot}.

\subsubsection{Archival host galaxy companion data}

The companion galaxy is covered by many of the same surveys mentioned in Sect. \ref{hostphot}. We use GALEX catalogue data, and measure photometry on PanSTARRS, VISTA, and WISE images with a 20\arcsec aperture. The galaxy is not detected in WISE W4. Similar to the host galaxy photometry, the data are corrected for Galactic extinction and given in AB magnitudes in Table \ref{table:hostphotcomp}. Additionally, the aperture contains a single clearly visible star, with red colours and tabulated magnitudes ranging from $g^\prime=22.58$ mag to $K_S=19.71$ mag (AB mags, corrected for extinction).We subtract the flux of this star to achieve our final values (it is not visible as a separate source in WISE).

\subsubsection{PMAS spectroscopy of the companion galaxy}

The companion galaxy, identified through {\hi} emission, and which showed an optical counterpart galaxy in catalogue images, was observed with the Potsdam
Multi-Aperture Spectrophotometer (PMAS) mounted on the 3.5~m telescope of Calar Alto Observatory \citep[Spain;][]{Roth2005}. PMAS is an integral field spectrograph composed of 16 × 16 square elements in the lens array. We used the 1\farcs0 spatial resolution configuration, which provides a field of view of 16$^{\prime\prime}\times16^{\prime\prime}$; the V500 grism was positioned at the grating position 143.5, which results in the wavelength range 3600-7500 Å with a spectral resolution of 1.55 Å (resolving power $R \sim600$). Four independent pointings were used to cover the emission of the complete galaxy. The data were reduced with the P3D data reduction tools \citep{Sandin2010}. In this paper we use the integrated spectrum obtained from the final data cube to search for emission lines to confirm the redshift derived from {\hi} emission. The analysis of the resolved observation is presented by \citet{Thoene2023}.

\section{Results}

\subsection{Broadband analysis of the host and companion}

A Spectral Energy Distribution (SED) analysis was performed for the host galaxy of GRB\,171205A using the photometry presented in Table~\ref{table:hostphot}. We used the \texttt{CIGALE}\footnote{\url{https://cigale.lam.fr/}} fitting code \citep{2005MNRAS.360.1413B, 2009A&A...507.1793N, 2019A&A...622A.103B} on its 2020 version. Here we highlight the main modelling choices and refer to \citep{2019A&A...622A.103B} for a detailed description. We consider a delayed star formation history where the star formation rate (SFR) is modelled using a simple exponential function normalised to the moment in which the SFR peaks. The age for the main stellar population in the galaxy is ranging from 1 Gyr to 12 Gyr and we consider an age for a late burst of 20 or 50 Myr. The choice of the Initial Mass Function (IMF) is the one described in \citet{2003PASP..115..763C} and a stellar population model using \citet{2003MNRAS.344.1000B}, assuming the metallicity, Z, with possible values of 0.008, 0.02 or 0.05.

\begin{figure}[h!]
\begin{center}
	 \centering 
	 \includegraphics[width=\columnwidth]{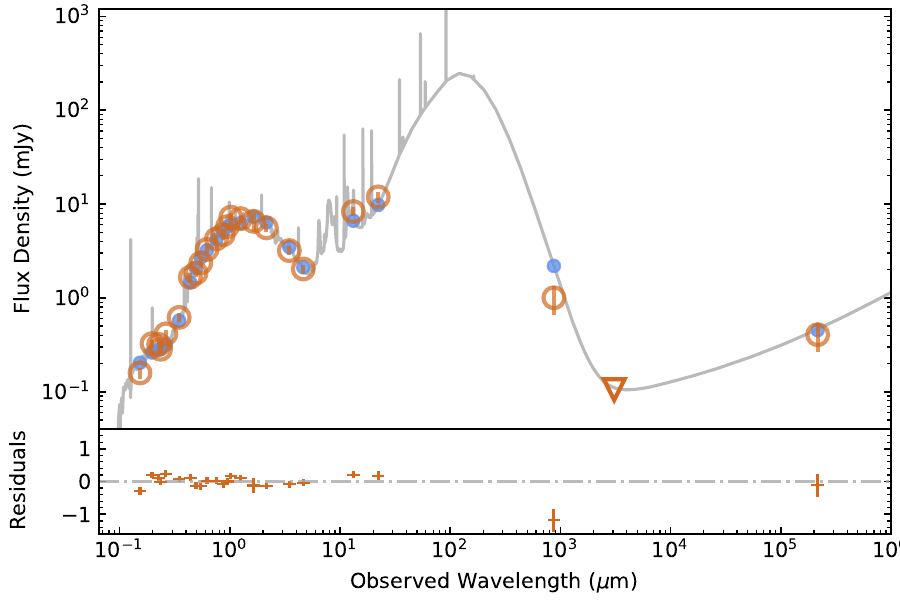} 
	 \caption{Best-fit SED model assuming a Milky Way extinction. The top panel shows the flux density distribution of the model together with the observed photometry from Table~\ref{table:hostphot}. Orange empty symbols are measured fluxes and upper limits, blue filled circles are the fluxes expected from the fitted model. The bottom panel shows the corresponding residuals for each photometric band.} 
	 \label{figure:best171205AMW}
\end{center}
\end{figure}

\begin{table}[h!]
\caption{Host galaxy properties as derived from the imaging and photometry presented in Table~\ref{table:hostphot} and corresponding best-fit SED model.}
\label{table:sed}
\centering
\begin{tabular}{c c}
\hline\hline
Property                        & Value                     \\
\hline 
Morphology                      & SbB                       \\
A$_V$ (mag)                     & $0.23\pm0.01$             \\
$\log_{10}(M/M_{\odot})$        & 10.29$_{-0.05}^{+0.06}$   \\
$\log_{10}({\rm SFR}/(M_{\odot}/{\rm yr}))$ & 0.09$_{-0.03}^{+0.03}$    \\
${\rm sSFR}\ ({\rm Gyr}^{-1})$              & 0.063 $\pm$ 0.010         \\
$\chi^{2}$/d.o.f.               & 2.1                       \\
\hline
\end{tabular}
\end{table}

\begin{table}[h!]
\caption{Companion galaxy properties derived from the SED fit of the photometry presented in Table~\ref{table:hostphotcomp}.}
\label{table:sed_companion}
\centering
\begin{tabular}{c c}
\hline\hline
Property                        & Value                     \\
\hline
Morphology                      & Irr                       \\
A$_V$ (mag)                     & $0.32 \pm 0.10$           \\
$\log_{10}(M/M_{\odot})$        & 9.10$_{-0.03}^{+0.02}$    \\
$\log_{10}({\rm SFR}/(M_{\odot}/{\rm yr}))$ & -2.02$_{-0.28}^{+0.17}$   \\
${\rm sSFR}\ ({\rm Gyr}^{-1})$              & 0.008 $\pm$ 0.004         \\
$\chi^{2}$/d.o.f.               & 0.98                      \\
\hline
\end{tabular}
\end{table}

To model the dust attenuation, we consider the modified \citet{2000ApJ...533..682C} attenuation law implemented in \texttt{CIGALE}. A Milky Way extinction law is adopted with R$_\textnormal{V}$ = 3.1 and a colour excess, for old and young populations, of the nebular lines ranging from 0.05 to 2.0 on 0.05 steps. The reduction factor for the colour excess is considered to be 0.44. We also let the slope for the attenuation curve to vary from -0.6 to 0.6 on 0.2 steps.
To estimate the re-emitted light on IR by dust, coming from stellar emission, we consider the \citet{2014ApJ...784...83D} models. Since Mid-infrared detections are available, we consider the slope for the dust mass heated by the radiation field intensity in heating intensity on dust to be $\alpha$ = 2, as found on \citep{Dale_2002}. The Active Galactic Nuclei (AGN) fraction is set to zero. 
In Fig.~\ref{figure:best171205AMW} we show the results for the SED fitting as well as the residuals for the measured fluxes. The only band that shows a significant deviation is the submillimetre observation, for which we assumed a multiplying factor to estimate the total flux density, that could be somewhat inaccurate. Due to this we increased the uncertainties of the data that uses this factor to 1/3 of the resulting flux density, to reduce their weight in the overall fit. The results of the fit are displayed in Table~\ref{table:sed}. A similar analysis was performed for the companion galaxy and the results are shown in Table~\ref{table:sed_companion}.

\subsection{CO absorption spectroscopy}

The ALMA observing windows of Band 3 were tuned to cover the range in which we would expect to observe the CO(1-0) molecular transition at the redshift of the GRB. The observations started on the 7th December, 2.02 days after the GRB, and covered the passage of the light curve peak, with flux densities reaching 60 mJy at 2.17 days after the burst. In spite of observing with the maximum possible continuum emission, and in spite of this being the second brightest afterglow observed at these wavelengths, there is no absorption detection. However, these observations can be used to determine a limit on the column density of CO along the line of sight of the GRB within its host galaxy. 

To estimate the molecular column density, in our previous publication \citep[see Sect. 3.3.5 from][]{deugarte2018} we used the formulation commonly used for quasar absorption systems \citep[e.g.,][]{Wiklind1995,Wiklind1997,Muller2011}.
This formulation assumes that the bright continuum source dominates over the background temperature ($T_c$) and the temperature of the foreground cloud ($T_{ex}$), as explained in Sect. 4.2 from \citet{martin2019}. A more careful analysis shows, however, that the continuum emission detected at the GRB positions is not luminous enough to fulfil this approximation.
In this work we used the radiative transfer fitting capabilities of MADCUBA package \citep{martin2019}, which takes as input for the radiative transfer solution the measured $T_c$ and the assumed $T_{ex}$ of the absorbing gas, to estimate a more accurate limit to the CO column density .
   
\begin{figure}[h!]
   \centering
   \includegraphics[width=\hsize]{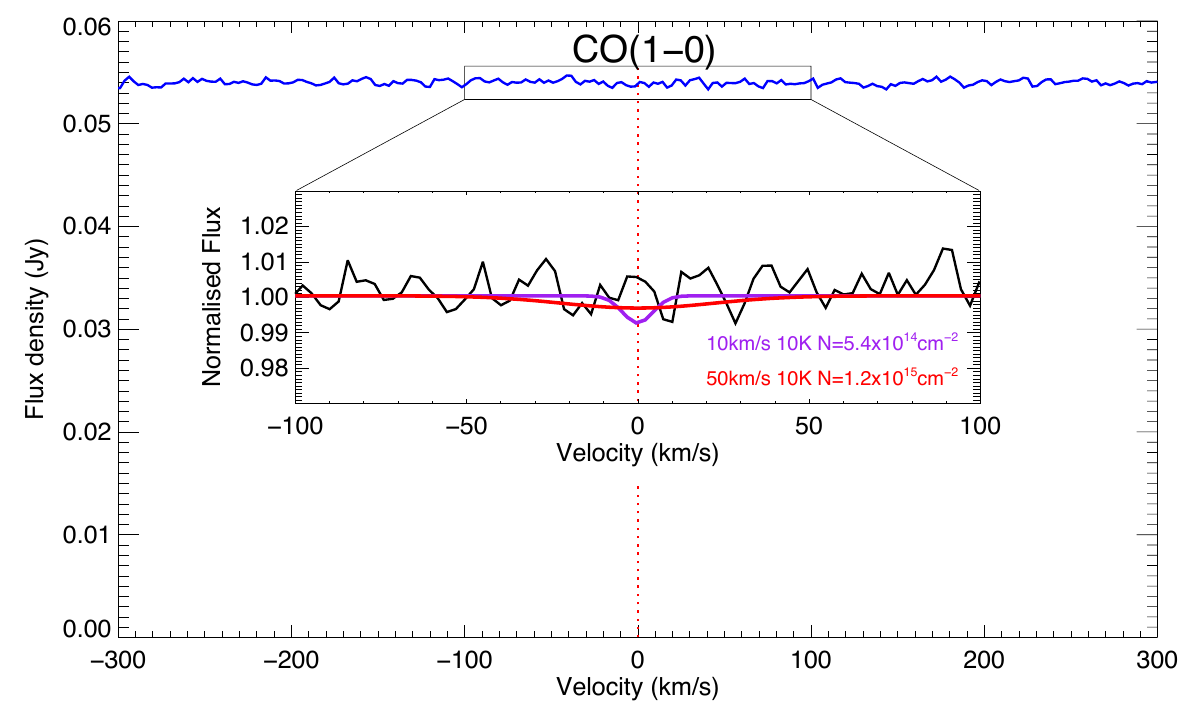}
      \caption{Afterglow spectrum centred at the frequency of CO(1-0). In spite of having a signal-to-noise ratio of over 200, no significant absorption is observed in this line of sight. The inset shows the profiles of CO(1-0) transitions with widths of 10 and 50 km/s which would have been at the limit of detection.
              }
         \label{fig:radiolc}
   \end{figure}

In Table~\ref{table:coabs} we present the limits to the column density based on the non detection of the CO $J=1-0$ transition against the measured background continuum emission. The Table presents the limits under the conservative assumption of a line width of 10 or 50~\kms, as well as for excitation temperature of the transition of 10, 20 and 50~K. The column densities in Table~\ref{table:coabs} correspond to the fit of MADCUBA to the $3~\sigma$ r.m.s. of the spectrum at the corresponding velocity resolution, depending on the assumed line width (i.e., the spectrum r.m.s. at 50~\kms will be a factor of $\sqrt{5}$ lower than that at 10~\kms). The fits for the $T_{ex}=10~\rm K$ are displayed in Fig.~\ref{fig:radiolc}.

\begin{table}[h!]
\caption{Derived limits to the foreground column density of CO.}
\label{table:coabs}
\centering
\begin{tabular}{c c c}
\hline\hline
T$_{ex} (\rm K)$    & Line width (km/s) & log~N (cm$^{-2}$) \\
\hline
10                  & 10                & < 14.73           \\
10                  & 50                & < 15.08           \\
20                  & 10                & < 15.30           \\
50                  & 10                & < 16.18           \\
\hline
\end{tabular}
\end{table}

The results in Table~\ref{table:coabs} show the strong dependence between the column density limit and the temperature of the foreground gas. At higher temperatures the assumption of $T_{ex}<T_c$ is broken and the absorption profile gets fainter independent of the column density, and therefore the column density is poorly constrained. 

This degeneracy can only be broken with ``simultaneous'' observations of multiple CO transitions. To illustrate that, we modelled with MADCUBA the expected line profiles for an assumed line width of $\Delta v=10$~\kms and a constant column density of $10^{15}~\rm cm^{-2}$, for the same three excitation temperatures assumed above. As background emission we used the one measured in this work and a spectral index of 0.5, therefore extrapolating the continuum emission as $F_\nu\propto\nu^{-0.5}$.
Additionally we also modelled the emission assuming a 3 times brighter GRB event illuminating the molecular gas. Peak modelled fluxes in both emission and absorption are presented in Table~\ref{table:comodel}.
This table shows how the temperature could be constrained within the achieved sensitivity if multiple transitions are observed, and how detectability improves with the brightness of the event, when we approach the condition of $T_{ex}<<T_c$ where absorption profiles are deeper. Future observations attempting to constrain the excitation temperature should thus focus on the first three CO transitions and can be best performed on events with very high luminosity and/or CO column density.

\begin{table}[]
    \centering
    \caption{Peak flux in mJy of modelled CO transitions for a flux density equivalent to the observation presented here (top) and three times more (bottom), assuming a spectrum with the shape F$_{\nu}\propto \nu^{-0.5}$.}
    \label{table:comodel}
    \begin{tabular}{l c c c}
    \hline
    \hline
         Transition & \multicolumn{3}{c}{$T_{ex}$}  \\
         &  10~K   &    20~K   & 50 K \\ 
    \hline
         \multicolumn{4}{c}{$F_\nu=F_{111~GHz}~\nu^{-0.5}$}  \\
         CO $J=1-0$ &   -0.86 &  -0.23  & -0.03 \\
         CO $J=2-1$ &   -1.22 &  -0.34  & 0.06  \\
         CO $J=3-2$ &   -0.42 &  0.13   & 0.58  \\
         CO $J=4-3$ &   -0.05 &  0.43   & 1.39  \\
    \hline
         \multicolumn{4}{c}{$F_\nu=3\times F_{111~GHz}~\nu^{-0.5}$}  \\
         CO $J=1-0$ &   -2.65 &  -0.78  & -0.13 \\
         CO $J=2-1$ &   -4.17 &  -1.65  & 0.23 \\
         CO $J=3-2$ &   -1.74 &  -0.98  & 0.20 \\
         CO $J=4-3$ &   -0.33 &  -0.15  & 1.04 \\
    \hline
    \end{tabular}
    \tablefoot{The model considers $N=10^{15}~\rm cm^{-2}$ and $\Delta v=10$~\kms . $F_{111~GHz}$ refers to the continuum flux measured in this work. Negative values correspond to lines observed in absorption. As a reference for detectability, in this work an r.m.s. of 0.2~mJy was achieved with 4.9~h of on source integration with ALMA.}
\end{table}

\subsection{CO emission throughout the galaxy}

   \begin{figure}
   \centering
   \includegraphics[width=\hsize]{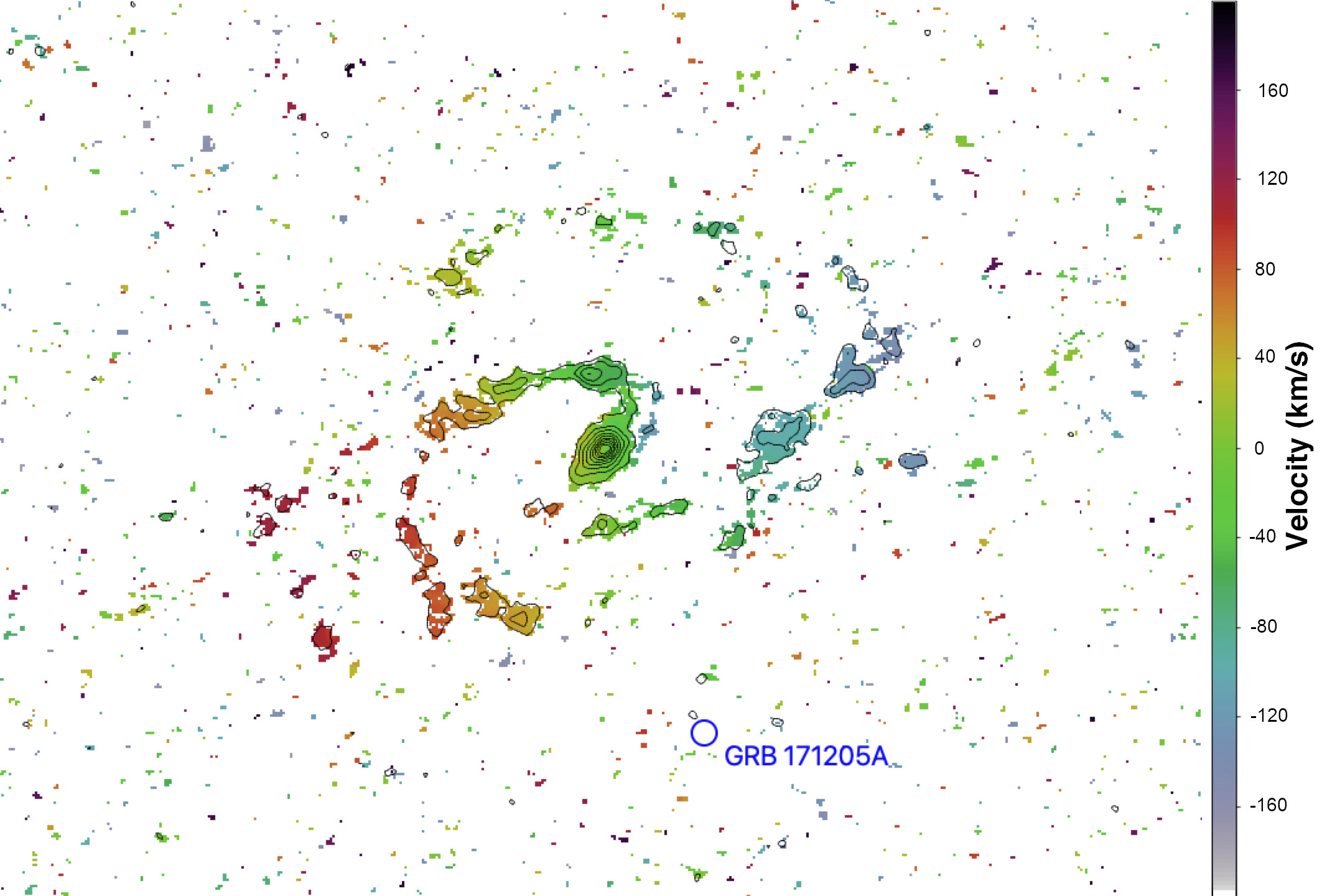}
   \includegraphics[width=\hsize]{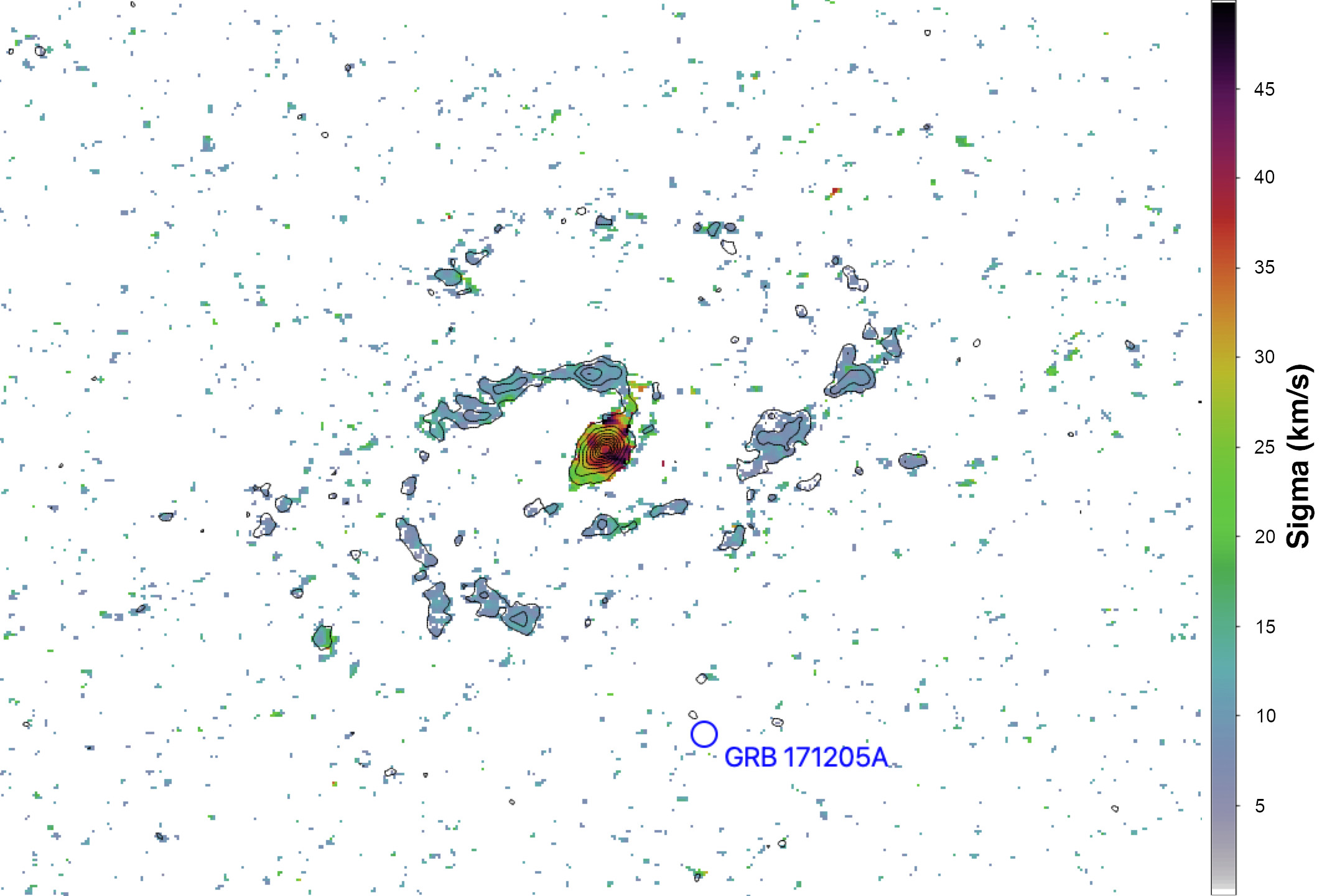}
      \caption{Velocity and width (sigma) of the emission line, as derived from a Gaussian fit.}
         \label{fig:almaco}
   \end{figure}

The observations performed at the CO($1-0$) transition, after continuum subtraction, show the host galaxy in emission at a spatial resolution equivalent to 147x191 pc. Using the ALMA data cubes we generated maps of flux, velocity, and velocity dispersion (sigma). To make these maps, we perform a Gaussian fit of the spectrum for each spaxel of the data cube. The flux map is calculated as the integral of each of these Gaussians. The velocity is calculated from the centroid of that Gaussian, as compared with the central spaxel of the galaxy, which corresponds to the following coordinates (J2000): R.A. 11:09:39.683, Dec: -12:35:11.74. Finally the line width is also determined as the sigma value (standard deviation) for each of these Gaussians. Since the signal to noise ratio in many of the spaxels was low, the maps required specific filtering to bring out the galaxy details. To this end, we only attempted the Gaussian fit in those spaxels in which the maximum spectral flux reached at least 2 times the r.m.s. of the image. Furthermore, we only accepted as good the fits in which the velocity was within 200 km/s from the central spaxel and where the fit parameters had individual errors smaller than 1.5 $\sigma$.

The CO flux map shows that one third of the CO emission is concentrated in the core and bar of the galaxy, whereas the rest of the emission extends through the spiral arms. As compared with the HST images the CO emission traces the core and spiral arms. In particular, the CO correlates with the dust lanes in the spiral arms.

   \begin{figure}
   \centering
   \includegraphics[width=0.8\hsize]{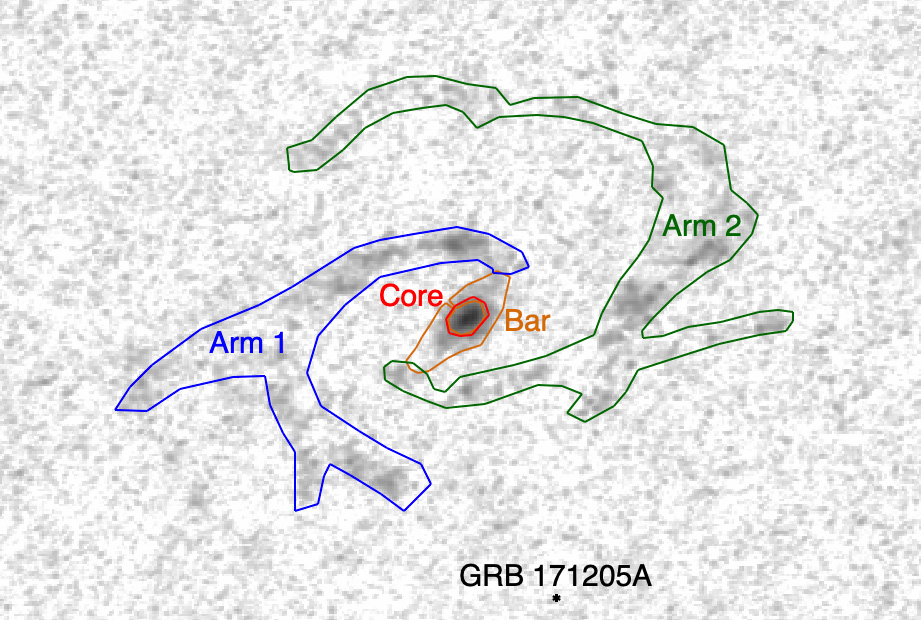}
   \includegraphics[width=\hsize]{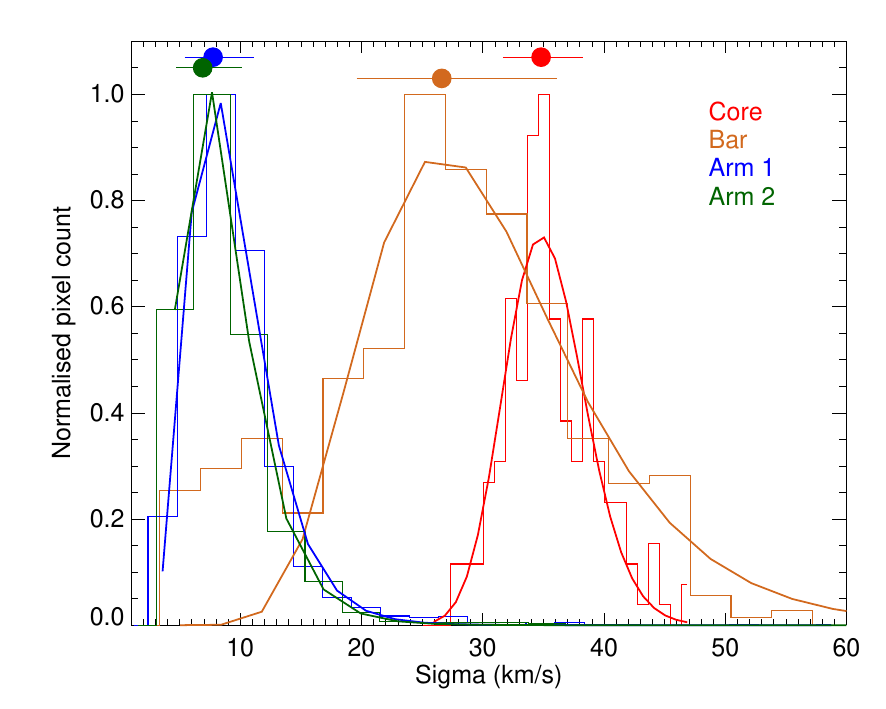}
      \caption{Top: Definition of the extracted regions within the galaxy. Bottom: Histograms with the distribution of CO line widths in the spaxels of the different galaxy regions.}
         \label{fig:sigmahist}
   \end{figure}
   
The velocity distribution of the CO emission follows the typical symmetric pattern of a spiral galaxy, with no apparent distortions. We have studied the CO line width in different regions of the galaxy, as defined in Fig.~\ref{fig:sigmahist}: The core, the bar, and its two main arms, which we identify as arm 1 and arm 2. The histograms show the distribution of line widths for the spaxels within each of those regions. Each of the histograms has been fit with a log-normal distribution. The broadest features are found at the core of the galaxy, with widths of $34.8_{-3.2}^{+3.5}$ km/s, the bar has a typical width of $26.6_{-7.0}^{+9.5}$ km/s, and both of the arms have consistent values of $7.8_{-2.3}^{+3.4}$ km/s and $6.9_{-2.2}^{+3.2}$ km/s, respectively. The measurements derived from the arms are consistent with the spectral resolution of the observations, and should be treated as upper limits. Again, the line widths behave according to the normal pattern of a spiral galaxy without indication of further distortions.

From the CO map we can measure the integrated flux of the host and from it derive the molecular mass and estimate the star formation rate, as shown in Table~\ref{table:co}. To allow easy comparison with the literature, we have used, for the calculation of the {\mhtwo} an $\alpha_{CO}=5.0$ M$_\odot$/(K km s$^{-1}$ pc$^{2}$). We also adopt a metallicity-dependent CO-to-H2 conversion factor based on \citet{Amorin2016}. Assuming a metallicity of 12+log(O/H) = 8.49 \citep{Thoene2023}, this results in $\alpha_{CO}=9.65$ M$_\odot$/(K km s$^{-1}$ pc$^{2}$).

   \begin{figure*}[h!]
   \centering
   \begin{tabular}{cc}
   \includegraphics[width=0.5\hsize]{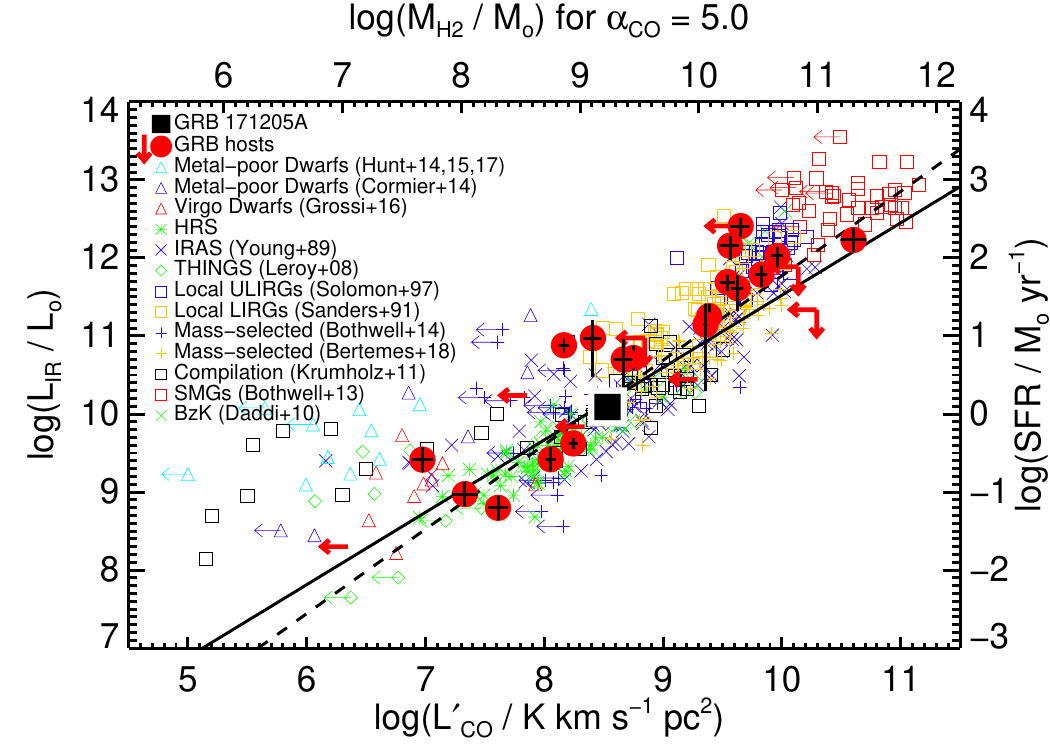}
   \includegraphics[width=0.46\hsize]{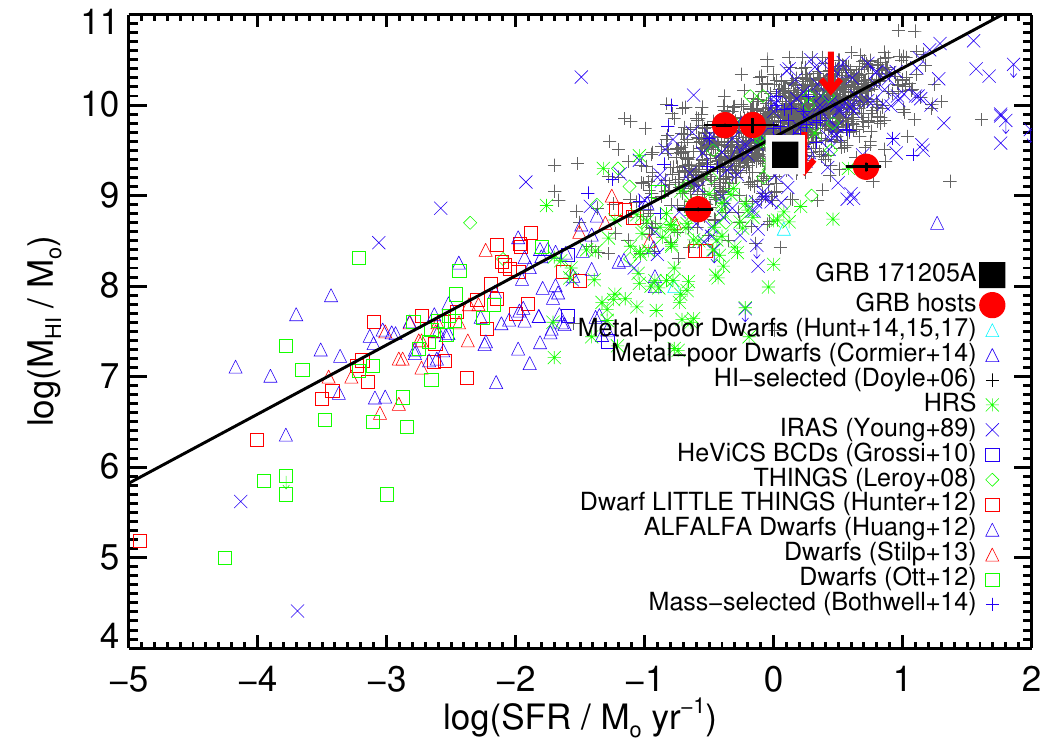} \\
   \end{tabular}
      \caption{Molecular (left) and atomic (right) gas masses as a function of SFR of hosts of GRBs (red circles) and other star-forming galaxies (other symbols, as noted in the legend). The host of {\grb} (large black square) has a factor of two lower atomic gas than the mean trend and a normal molecular gas mass for its SFR. The Figure is adopted from \citet{Michalowski2018co} and  \citet{Michalowski2015}. The CO measurements for the GRB hosts are from \citet{Michalowski2018co}, \citet{Hatsukade2020}, \citet{deugarte2020}, and \citet{Heintz2021}.}
         \label{fig:sfrgas}
   \end{figure*}

The SFR measured with the SED modelling (Table~\ref{table:sed}) implies an expected value of
$\log(\mhtwo/\msun)=9.12$ and $9.33$ from the galaxy-integrated integrated Kennicutt-Schmidt law of \citet[][eq.~1]{Michalowski2018co} and \citep[][Fig.~7]{Carilli2013}. The measured molecular gas mass of
$\log(\mhtwo/\msun)=9.25$ (Table~\ref{table:co}) is consistent with these values, so this galaxy has molecular gas mass similar to what is found for other galaxies with similar SFRs.

We show these measured SFRs and {\mhtwo} for the GRB hosts in Fig.~\ref{fig:sfrgas} (the figure adopted from \citealt{Michalowski2018co}), including GRB hosts observed at CO by \citet{Hatsukade2020}, \citet{deugarte2020}, and \citet{Heintz2021}. 
As calculated above, the {\grb} host is consistent with
the typical relation for star-forming galaxies.
We compare the GRB hosts to other galaxy samples, as compiled in \citet{Michalowski2015} and \citet{Michalowski2018co}\footnote{\citet{young89b}, \citet{sanders91}, \citet{solomon97}, \citet{doyle06}, \citet{leroy08}, \citet{boselli10,boselli14}, \citet{catinella10}, \citet{daddi10},   \citet{grossi10}, \citet{michalowski10smg}, \citet{magdis11}, \citet{cortese12b,cortese14}, \citet{hunter12}, \citet{huang12}, \citet{magnelli12b}, \citet{ott12}, \citet{bothwell13,bothwell14}, \citet{stilp13}, \citet{wang13}, \citet{ciesla14}, \citet{cormier14}, \citet{hunt14b}.}. 
This compilation includes local spirals, dwarfs, and (ultra)luminous infrared galaxies, as well as samples selected with {\hi} or stellar mass.

Giant molecular clouds (GMCs) are prime locations for star formation. As such, the progenitors of GRBs could be linked to such clouds. Our CO observations are well suited to search for such a cloud linked to \grb, with a beam size comparable to the size of GMCs. However, our CO data shows no detection at the site of the GRB. In Table~\ref{table:co} we include limits for the brightness, luminosity, molecular mass, and SFR. These limits are within the range of GMCs, and argue against the association of {\grb} with such an environment. Even more interesting is the fact that our SFR limit at the location of the GRB (SFR$_{GRB site}<2.4\times10^{-4}$ M$_\odot$ yr$^{-1}$) is significantly lower (over 80 times) than the measurement obtained from the optical emission at the GRB site from optical data by \citet[][ $0.020\pm0.001$ M$_\odot$ yr$^{-1}$]{Thoene2023}. This is shows that the star formation that produced this GRB was either not triggered by molecular gas, or that the molecular gas was later fully ionised by radiation from the young stars or from the GRB itself.

\begin{table}
\caption{Host galaxy and GRB location properties derived from the CO(1-0) emission. For the calculation of the mass of H$_2$ we are assuming both a generic conversion factor of $\alpha_{CO}=5.0$ M$_\odot$/(K km s$^{-1}$ pc$^{2}$) and a metallicity dependent value of $\alpha_{CO}=9.65 $ M$_\odot$/(K km s$^{-1}$ pc$^{2}$. The star formation rate is calculated based on \citet{Gao2004}.}
\label{table:co}
\centering
\begin{tabular}{c c}
\hline\hline
\multicolumn{2}{c}{\textbf{Host galaxy}}\\
\hline
Property                                        & Value                     \\
\hline
F$_{int}$ ( Jy~km~s$^{-1}$)                     & $5.43\pm0.06$             \\
$L'_{CO}$ (K~km~s$^{-1}$~pc$^2$)                & $(3.40\pm0.04)\times10^8$ \\
\mhtwo (M$_\odot$) [$\alpha_{CO}=5.0$]          & $(1.70\pm0.02)\times10^9$ \\
SFR (M$_\odot$~yr$^{-1}$) [$\alpha_{CO}=5.0$]   & $2.38\pm0.26$             \\
\mhtwo (M$_\odot$) [$\alpha_{CO}=9.65$]         & $(3.44\pm0.04)\times10^9$ \\
SFR (M$_\odot$~yr$^{-1}$) [$\alpha_{CO}=9.65$]  & $4.59\pm0.50$             \\
\hline\hline
\multicolumn{2}{c}{\textbf{GRB site}}\\
\hline
Property                                        & Value                     \\
\hline
F$_{int}$ ( Jy~km~s$^{-1}$)                     & $<5.6\times10^{-4}$             \\
$L'_{CO}$ (K~km~s$^{-1}$~pc$^2$)                & $<3.5\times10^4$ \\
\mhtwo (M$_\odot$) [$\alpha_{CO}=5.0$]          & $<1.8\times10^5$ \\
SFR (M$_\odot$~yr$^{-1}$) [$\alpha_{CO}=5.0$]   & $<2.4\times10^{-4}$             \\
\mhtwo (M$_\odot$) [$\alpha_{CO}=9.65$]         & $<3.4\times10^5$ \\
SFR (M$_\odot$~yr$^{-1}$) [$\alpha_{CO}=9.65$]  & $<4.8\times10^{-4}$             \\
\hline
\end{tabular}
\end{table}

\subsection{HI emission of the host}

The {\hi} datacube reveals a clear detection of the {\grb} host.
Fig.~\ref{figure:almahost} shows the intensity map of the {\hi} line overlaid on optical data from HST, whereas Fig.~\ref{fig:hilarge} has a larger filed using PanSTARRS image as background. Fig.~\ref{fig:hispec} presents the {\hi} spectrum, whereas Table~\ref{tab:mhi} lists the properties derived from the {\hi} data. We calculated redshifts from the emission-weighted frequency of the {\hi} line, integrated fluxes within the dotted lines on Fig.~\ref{fig:hispec} (placed at frequencies at which the emission of both peaks drops to zero). The {\hi} line luminosities were calculated using equation 3 in \citet{solomon97}, and the neutral hydrogen masses using equation 2 in \citet{devereux90}.

The {\hi} moment 0 map of the {\grb} host reveals a gas distribution that departs from what we see in the optical and molecular gas. In the case of {\hi} there are two blobs of gas corresponding to the outer part of the disk, along its major axis with a lack of gas towards the core of the galaxy. The overall gas distribution is slightly more prominent towards the southwest of the galaxy where we find the GRB. The rotation of the gas, as shown by \citep{Thoene2023} is similar to the rest of the disk as compared to their MUSE/VLT (stellar emission) and our ALMA (molecular gas) data. The {\hi} could correspond to a toroidal distribution of the atomic gas, with a lower concentration in the core of the galaxy. The fact that the velocity distribution is similar to the rest of the disk argues against a major galaxy interaction in the past. We do note, however, that the overall emission of {hi} is displaced, or extended, towards the southwest, where the GRB is located, which could be sign of a minor interaction.

The SFR measured from the SED modelling (Table\ref{table:sed}) implied the expected $\log(\mhi/\msun)=9.71$ \citep[][eq.~1]{Michalowski2015}. 
The measured atomic gas mass of $\log(\mhi/\msun)=9.45\pm0.05$ (Table~\ref{tab:mhi}, consistent with the value of 9.49$\pm$0.04 independently determined by \citealt{arabsalmani2022})
is 0.26\,dex lower. The scatter in this relation is of that order, so the {\grb} host is slightly deficient in atomic gas (but within the scatter), or its SFR is enhanced by a factor of two.

As for atomic gas, we show the molecular gas mass as a function of SFR in Fig.~\ref{fig:sfrgas}. The {\grb} host is lower by a factor of two than the average relation, within the scatter of other galaxies.

The molecular gas fraction [$\mhtwo/(\mhtwo+\mhi)$] of the {\grb} host is 39\%. This is a high but typical value for star-forming galaxies \citep{Young1989,Devereux1990,Leroy2008,Saintonge2011,Cortese2014,Boselli2014,Galbany2017,michalowski18sn}.

\subsection{Host galaxy environment}

\begin{figure*}[h!]
\centering
\includegraphics[width=0.9\hsize]{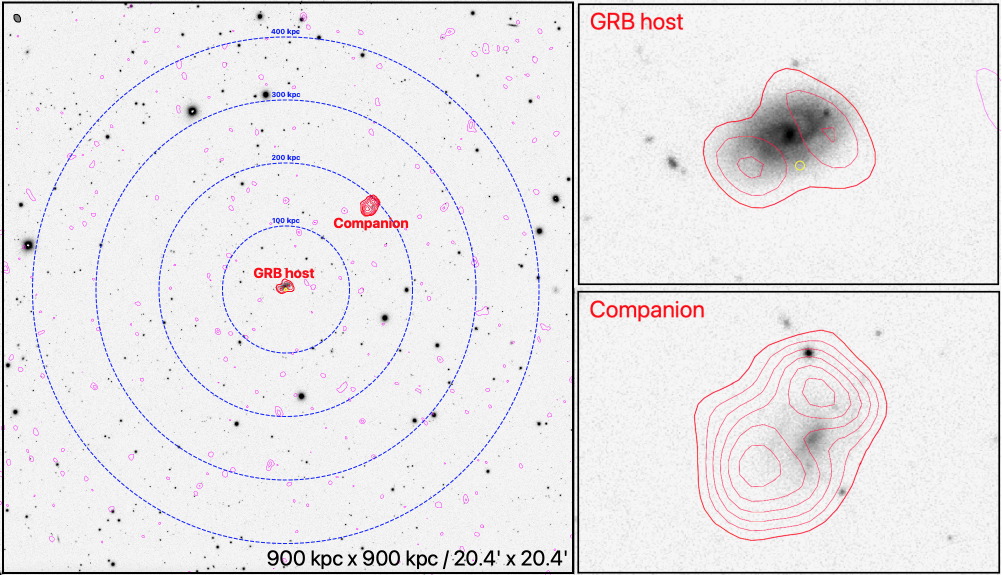}
    \caption{Identification of the {\hi} emitting regions within the GMRT field of view surrounding GRB\,171205A in an analysis optimised for wide field. The background image was obtained from the PanSTARRS catalogue, combining all the filters ($g^\prime r^\prime i^\prime z^\prime y$). 
    The only two objects significantly detected both in the GMRT and the JVLA data (GRB host and companion) are marked in red. The field of view covers a projected area of $900\times900$ kpc.}
        \label{fig:hilarge}
\end{figure*}

\begin{figure*}[h!]
\begin{center}
\begin{tabular}{cc}
\includegraphics[width=8cm,clip]{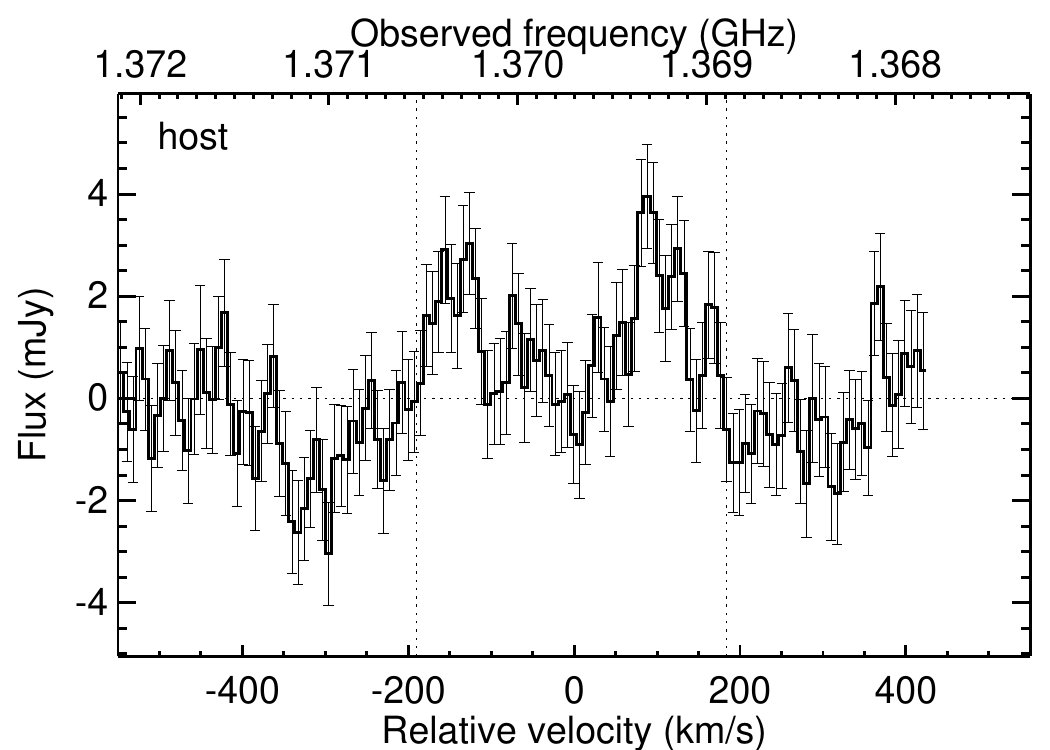} & 
\includegraphics[width=8cm,clip]{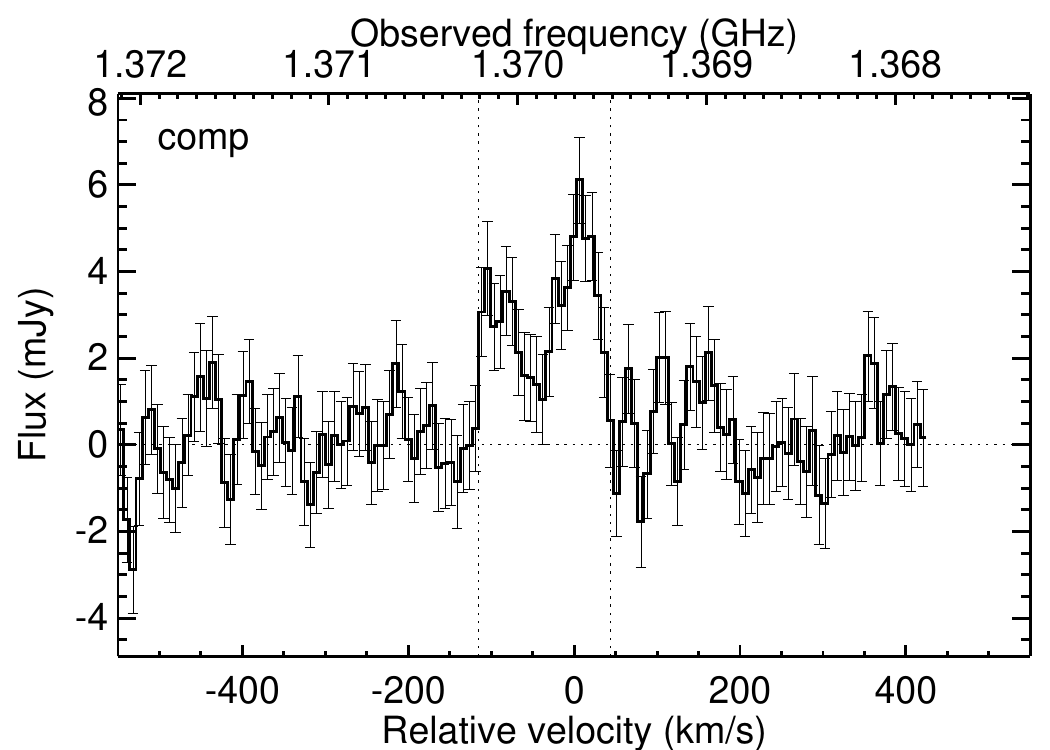} \\
\end{tabular}
\end{center}
\caption{{\hi} spectra of the {\grb} host and its companion galaxy. The line fluxes were calculated by integration of the line within the vertical dotted lines. The velocities are relative to the velocity of the central pixel in the CO cube with, i.e. $z = 0.03702$.
}
\label{fig:hispec}
\end{figure*}

\begin{table*}
\small
\centering
\caption{{\hi} properties of the host and confirmed companion.\label{tab:mhi}}
\medskip
\begin{tabular}{lcccccccc}
\hline
Object & R.A. & Dec. & \multicolumn{2}{c}{Distance to {\grb}} & \zhi & $F_{\rm int}$   &  $\log(\lphi)$          & $\log(\mhi)$ \\
    & (deg) & (deg) & ($\arcsec$) & (kpc) &      & (Jy km s$^{-1}$) & ($\mbox{K km s}^{-1} \mbox{ pc}^2$) & (${\rm M}_\odot$) \\
(1)    & (2)     & (3)  & (4)           & (5) & (6) & (7) & (8) & (9)  \\
\hline
host & $167.415692$ & $-12.589399$ & $\cdots$ & $\cdots$ & $0.037018 \pm        0.000045$ & $0.45 \pm 0.05$ & $11.270 \pm 0.050$ & $9.451 \pm 0.050$  \\
comp & $167.362972$ & $-12.539701$ & $258$ & $188$ & $0.036916 \pm        0.000009$ & $0.49 \pm 0.03$ & $11.304 \pm 0.030$ & $9.485 \pm 0.030$  \\
\hline
\end{tabular}
\tablefoot{
(1) Object. (2,3) Position. (4,5) Distance to the {\grb} host in arcsec and kpc. (6) Redshift determined from the emission-weighted frequency of the {\hi} line. (7) Integrated flux within the dotted lines on Fig.~\ref{fig:hispec}. (8) {\hi} line luminosity using equation 3 in \citet{solomon97}. (9) Neutral hydrogen mass  using equation 2 in \citet{devereux90}.
}
\end{table*}

\begin{figure}
\centering
\includegraphics[width=\hsize]{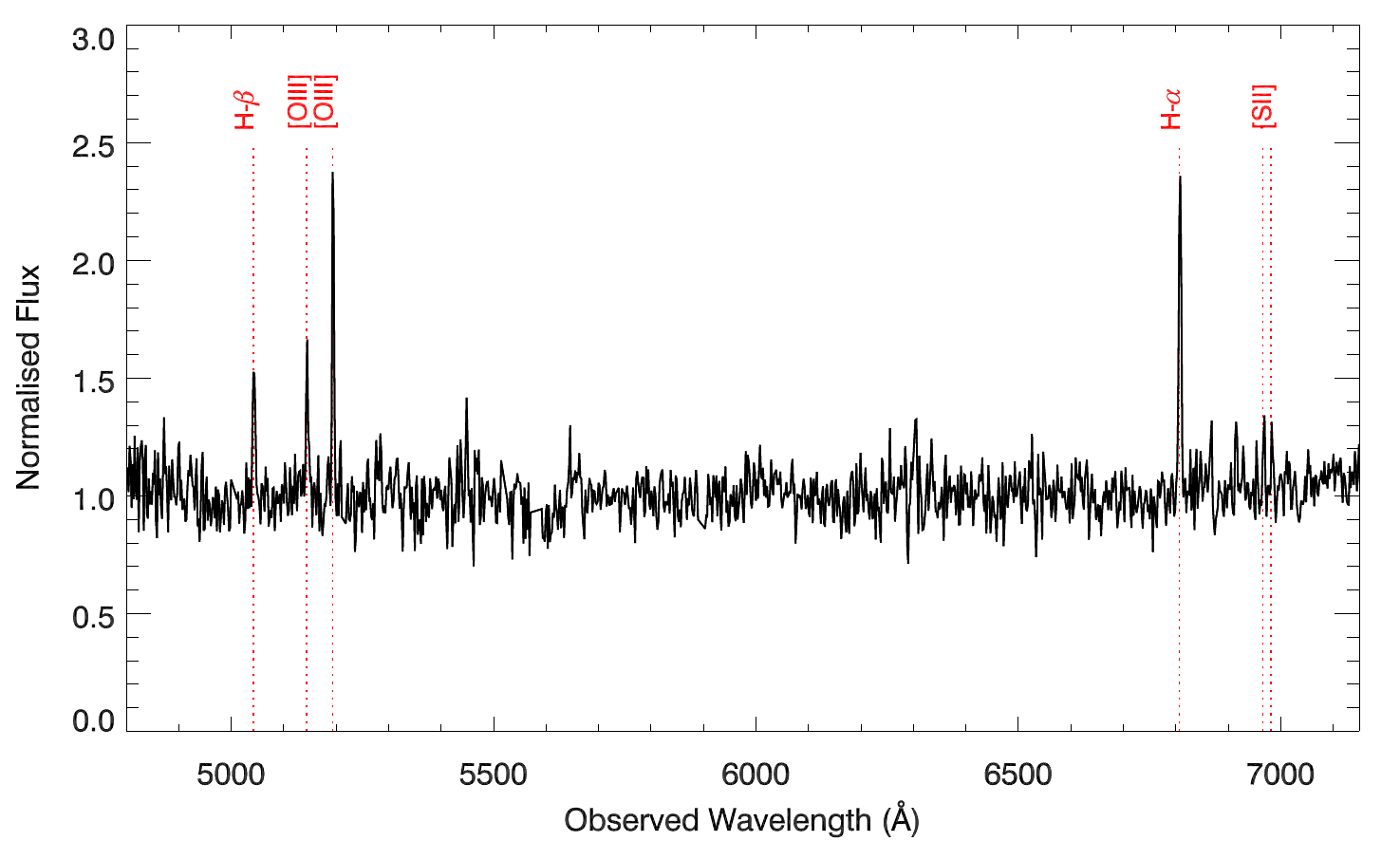}
    \caption{Optical spectrum of the companion galaxy, showing emission features at a redshift of $z=0.0369$.}
        \label{fig:comp}
\end{figure}

In the {\hi} data cube we detect a companion irregular galaxy at 188\,kpc in projection from the {\grb} host with the spectra at overlapping velocities and the central velocity difference of only 30 km\,s$^{-1}$ (see Fig.\ref{fig:hilarge} for the map, Fig.~\ref{fig:hispec} for the spectrum and Table ~\ref{tab:mhi} for derived properties). The centre of the {\hi} distribution of this galaxy is shifted from the optical galaxy centre. The redshift of the galaxy was confirmed through the identification of multiple emission lines in the optical spectroscopy performed with PMAS\footnote{Project no. 19A-3.5-029, PI: L. Izzo}. Using the integrated spectrum of the galaxy (see Fig.\ref{fig:comp}), we detect emission of H$\alpha$, H$\beta$, and [OIII] at a redshift of 0.0369, consistent with the one of the host galaxy.

On the integrated {\hi} map we identified several faint objects with lower significance. However, when comparing them with the observations obtained with JVLA  none of them were confirmed, and we interpret them as noise fluctuations. The companion galaxy has a similar atomic gas mass to the one of the {\grb} host. However, if we compare the stellar masses derived from the SED fit, the host galaxy of GRB is 12 times more massive than the companion.

\section{Discussion}

\begin{figure}[h!]
\centering
\includegraphics[width=\hsize]{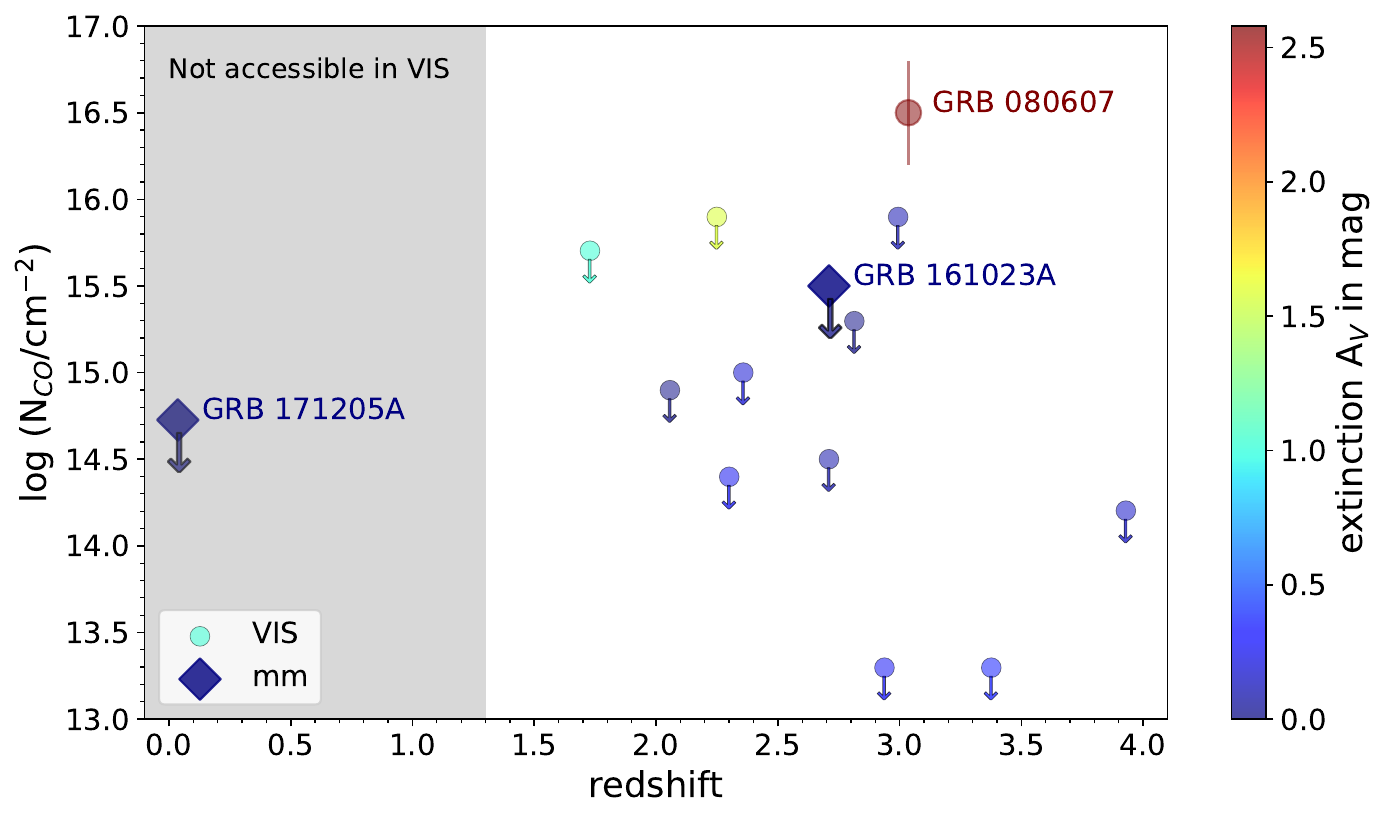}
    \caption{Comparison of observations of CO in GRB sight-lines using optical spectroscopy \citep[][]{Prochaska2009,Bolmer2019,Heintz2019} and millimetre observations \citep{deugarte2018}. There is currently only one detection, GRB\,080607, obtained with optical spectroscopy \citep[]{Prochaska2009}. Millimetre observations allow to expand the search to all redshifts. The datapoints are colour coded according to the host galaxy line of sight extinction \citep[][]{Prochaska2009,Heintz2019,Izzo19Nat}.}
        \label{fig:co_z}
\end{figure}

The non-detection of the CO($1-0$) absorption feature and the implied limit on the CO abundance at the location of the GRB is consistent with the overall dearth of CO absorption in GRB sight lines \citep{Bolmer2019,Heintz2019}. To date, CO absorption has only been detected in the single case of GRB\,080607 \citep{Prochaska2009}, which is also situated in one of the dustiest and metal-rich GRB environments observed. In that case the column density was measured to be log(CO/cm$^{-2}$) = 16.5$\pm$0.3. These constraints, together with the observed H$_2$ and C\,{\sc i} column densities and overall low molecular gas fractions, are still consistent with the observed properties of diffuse or translucent molecular clouds in the Milky Way \citep{Burgh2010}. 
The rest-frame UV observations require redshifts $z\gtrsim 1.3$ to be detectable from the ground (AX bandheads are found at $\approx$1500 {\AA}). 
However, targeting CO in absorption at millimetre wavelengths over the typical rest-frame UV 
has the advantage that this can be done at any redshift, including nearby GRB sight lines (see Fig.~\ref{fig:co_z}). 
Furthermore, the millimetre GRB afterglow reaches peak light later and maintains its brightness for longer periods of time. We note that successful searches for molecular absorption will require both bright afterglows and dusty environments, in which the molecules can subsist.

Our data show three main aspects of the host of {\grb}. First, it has a regular CO velocity field (which is similar to the H$\alpha$ velocity field; \citealt{Thoene2023}) but no CO emission at the GRB location. Second, the {\hi} gas does not follow a uniform distribution and is instead found in two main emission blobs along the major axis of the projected galaxy disk, with a slight asymmetry and  shift towards the southwest, where the GRB is located. We note that the JVLA reduction that we use does not show the clump mentioned by \citep{arabsalmani2022} as the possible remnant of a recent interaction. This is also not seen in our GMRT observation, indicating that it is likely an artefact of their data reduction. Third, it has a similar molecular gas and a factor of two lower atomic gas mass to other star forming-galaxies of similar SFRs.

The optical and CO emissions argue against any major galaxy interaction in the recent past, since there is no trace of distortion in them. The {\hi} images do show an altered atomic gas field. This {\hi} field reaches larger distances from the galaxy core and is less gravitationally bound to it, so it is more likely to be affected by a weak interaction or by the inflow of a pristine gas cloud. A distortion of the {\hi} field could have led to gas compression and to the formation of new stars. One of such stars could be the GRB progenitor.

It has been shown that ongoing interactions with other galaxies can lead to an increase in SFRs \citep[as for NGC\,2770, the SN factory][]{thone09,michalowski20b}. This is not supported by the regular CO and H$\alpha$ velocity field that this galaxy has, but we could consider if the interaction with the companion 188\,kpc away could be responsible for the enhanced SFR.
This distance is too high for the {\it current} interaction to lead to SFR enhancement, but the interaction may have been stronger in the past.
Assuming a velocity between the two galaxies to be 200-400 km\,s$^{-1}$ (similar to the higher end of velocities seen in the Local Group, \citealt{vandenbergh1999}), at this distance from the host a close encounter between the two galaxies would have taken place at least 400-800 Myr in the past. 
The small relative radial velocity between the two galaxies indicates that they are most probably moving close to the plane perpendicular to the line of sight and the projected distance is very similar to their real distance. Both galaxies show an {\hi} distribution displaced with respect to their stellar light, which argues in favour of the interaction scenario, even if it only a minor one that limits its effect to the {\hi} field.

Numerical simulations show that SFRs of interacting galaxies may be enhanced by a factor of a few for several hundred of Myr after the first pericentre passage, even when galaxies move away from each other by 100\,kpc or more \citep{dimatteo08,moreno21}.
At the apocentre passage the interacting galaxies keep their spiral morphology (the host is a spiral). 
Hence, while we conclude that SFR of the host could have been enhanced by the interaction with the companion if they have been at the pericentre a few hundred Myr ago, the regular velocity field does not support this scenario.

Similarly, the global properties of the galaxy do not support the scenario of a completed or ongoing merger with a dwarf galaxy. In such a case the atomic gas could be expelled in form of tidal tails, but the SFR would be enhanced and the velocity field as traced by CO and H$\alpha$ would also be disturbed, which is not what we observe.

The asymmetry of the {\hi} distribution and the extension to the southwest are similar to those found in other long GRB hosts  detected at {\hi}, namely for GRB\,980425 \citep{Michalowski2015,Michalowski2016,arabsalmani15b,arabsalmani19} and GRB\,060505 \citep{Michalowski2015}. Broad-line Ic supernovae seem to show a similar asymmetry (SN\,2009bb, \citealt{michalowski18sn} and SN\,2002ap, \citealt{michalowski20}) and also the hosts of the elusive fast radio bursts might exhibit strong {\hi} line asymmetry \citep{michalowski21}.
Hence, the {\grb} host is the fifth example of an exploding massive star which was born close to an asymmetric {\hi} concentration. This strengthens the hypothesis put forward in \citet{Michalowski2015} that these massive stars are born as a result of a recent atomic gas inflow from the intergalactic medium.

The only counterexamples are the hosts of the SN-less GRB\,111005A and the peculiar transient AT\,2018cow, likely not associated to the GRB phenomena. There is no strong concentration close to the position of these transients  \citep{Lesniewska2022,Michalowski18grb,roychowdhury19}. However, the nature of these two objects is still highly debated \citep{prentice18,liu18,riverasandoval18,fox19,huang19,kuin19,margutti19,morokumamatsui19,perley19,soker19,lyutikov19,bietenholz20,lyman20,Michalowski18grb}.

\section{Conclusions}
We present a deep observation of {\grb} obtained with ALMA, with the initial goal of performing absorption spectroscopy. Most of the data were collected 48 hrs after the burst, while the afterglow had a flux density of $\sim 60$ mJy. This results in a SNR of over 200 per resolution element at a spectral resolution of 10 km/s. In spite of the high SNR there is no detection of CO($1-0$) in absorption down to a column density of $\sim10^{15}$ cm$^{-2}$, depending of the assumed line width and temperature. This limit is comparable to the ones usually obtained with ground-based optical spectroscopy and 1.5 dex stronger than the only existing detection for a GRB in optical spectroscopy, which was associated to a very dusty sight line. Millimetre observations have the advantage of allowing measurements at low redshift, but, like optical spectroscopy, they require a dusty environment for CO to be detectable with current instrumentation. In future studies, simultaneously observing different transitions on a bright event could help to constrain the temperature and the overall column density. This will be also useful in emission to better constrain the properties of the molecular gas within the host.

After continuum subtraction, our data delivered a highly resolved study of a GRB host galaxy in CO, at a spatial resolution of $228\times 176$ pc. We combined these data with resolved observations of the galaxy in {\hi} and with a multi-wavelength SED from which we can derive the integrated properties of the host. The CO emission traces the core, bar and dust lanes along the arms of the spiral galaxy. The velocity and velocity dispersion patterns correspond to an undisturbed galaxy. There is no emission detected at the location of the burst. At our resolution and sensitivity we could have been able to detect the emission of a giant molecular cloud of $\sim10^5$ M$_{\odot}$.

The {\hi} emission departs from a uniform disk and can be described as two asymmetric blobs with little or no emission within the galaxy core, and a slight excess towards the southwest of the host galaxy, in the direction where the GRB is located. With our GMRT data and a reanalysis of the JVLA data, we cannot confirm the clump reported by \citet{arabsalmani2022}, which they interpreted as an interaction remnant. This is also not seen at any other wavelength and we consider it an artefact of their reduction. The asymmetry and centroid displacement could be due to some sort of minor galactic interaction dominated by atomic gas. We searched for possible interacting galaxies and only found one convincing candidate, with both {\hi} emission and optical counterpart, at 188 kpc of the GRB host. This distance seems large, but this other galaxy is {\hi}-rich and also shows a displacement in its {\hi} field. Hence, a past mild interaction of this relatively distant galaxy pair could be considered as a plausible cause for the distortion in the {\hi} fields and ultimately for the star formation giving rise to {\grb}. This is a speculative conclusion but supported by the statistical existence of several other cases of GRB hosts with distorted {\hi} fields.

The lack of CO in absorption and emission argues against the preferential formation of GRB progenitors in molecular clouds, which is in agreement with previous reports of an overall dearth of CO in GRB sight lines. An alternative scenario would be that radiation from the young stars or even the GRB itself are fully ionising the molecular gas, leaving no detectable CO at the time of our observation. On the other hand, the distorted {\hi} field may be seen as a possible trigger for the star formation that gave rise to the GRB progenitor, which would favour the theories that link GRBs to metal poor atomic gas.

\begin{acknowledgements}

M.J.M. and A.L.~acknowledge the support of the National Science Centre, Poland through the SONATA BIS grant 2018/30/E/ST9/00208.
This research was funded in whole or in part by National Science Centre, Poland (grant number: 2021/41/N/ST9/02662).
For the purpose of Open Access, the author has applied a CC-BY public copyright licence to any Author Accepted Manuscript (AAM) version arising from this submission.
MPK acknowledges support from the First TEAM grant of the Foundation for Polish Science No. POIR.04.04.00-00-5D21/18-00.

This paper makes use of the following ALMA data: ADS/JAO.ALMA\#2017.1.01695.T, ADS/JAO.ALMA\#2017.1.01526.T, and ADS/JAO.ALMA\#2018.1.01635.S. ALMA is a partnership of ESO (representing its member states), NSF (USA) and NINS (Japan), together with NRC (Canada), MOST and ASIAA (Taiwan), and KASI (Republic of Korea), in cooperation with the Republic of Chile. The Joint ALMA Observatory is operated by ESO, AUI/NRAO and NAOJ.

We thank the staff of the GMRT that made these observations possible. GMRT is run by the National Centre for Radio Astrophysics of the Tata Institute of Fundamental Research.

Partly based on observations made with the Gran Telescopio Canarias (GTC), installed in the Spanish Observatorio del Roque de los Muchachos of the Instituto de Astrofísica de Canarias, in the island of La Palma.

AIPS is produced and maintained by the National Radio Astronomy Observatory, a facility of the National Science Foundation operated under cooperative agreement by Associated Universities, Inc. 

This article has been supported by the Polish National
Agency for Academic Exchange under Grant No. PPI/APM/2018/1/00036/U/001.
\end{acknowledgements}

\bibliographystyle{aa}
\bibliography{171205A_bib}

\end{document}